\documentclass[seceq]{ptptex}

\usepackage{epsf}
\usepackage{graphicx}
\makeatletter

\newcommand{\beq}{\begin{equation}}
\newcommand{\eeq}{\end{equation}}
\newcommand{\beqn}{\begin{eqnarray}}
\newcommand{\eeqn}{\end{eqnarray}}

\newcommand{\pa}{\partial}

\newcommand{\varep}{\varepsilon}

\def\agt{\mathrel{\raise.3ex\hbox{$>$}\mkern-14mu\lower0.6ex\hbox{$\sim$}}}
\def\alt{\mathrel{\raise.3ex\hbox{$<$}\mkern-14mu\lower0.6ex\hbox{$\sim$}}}
\def\hbar{\mathrel{\raise.7ex\hbox{--}\mkern-14mu\hbox{$~h$}}}

\title{%        %You can use \\ for explicit line-break
Radiation Magnetohydrodynamics for 
Black Hole-Torus System in Full General Relativity: 
A Step toward Physical Simulation
}

\author{%       %Use \scshape  for the family name
Masaru \textsc{Shibata} and Yuichiro \textsc{Sekiguchi}
}

\inst{%     %Affiliation, neglected when [addenda] or [errata]
Yukawa Institute for Theoretical Physics, Kyoto University, Kyoto
606-8502, Japan 
}

\abst{A radiation-magnetohydrodynamic simulation for the black
hole-torus system is performed in the framework of full general
relativity for the first time.  A truncated moment formalism is
employed for a general relativistic neutrino radiation transport.
Several systems in which the black hole mass is $M_{\rm BH}=3$ or
$6M_{\odot}$, the black hole spin is zero, and the torus mass is
$\approx 0.14$--$0.38M_{\odot}$ are evolved as models of the remnant
formed after the merger of binary neutron stars or black hole-neutron
star binaries. The equation of state and microphysics for the
high-density and high-temperature matter are phenomenologically taken
into account in a semi-quantitative manner. It is found that the temperature in
the inner region of the torus reaches $\agt 10$~MeV which enhances a
high luminosity of neutrinos $\sim 10^{51}$~ergs/s for $M_{\rm
BH}=6M_{\odot}$ and $\sim 10^{52}$~ergs/s for $M_{\rm BH}=3M_{\odot}$.
It is shown that neutrinos are likely to be emitted primarily toward
the outward direction in the vicinity of the rotational axis and their energy
density may be high enough to launch a low-energy short gamma-ray
burst via the neutrino-antineutrino pair-annihilation process with the
total energy deposition $\sim 10^{47}$--$10^{49}$~ergs. It is also
shown in our model that for $M_{\rm BH}=3M_{\odot}$, the neutrino luminosity is
larger than the electromagnetic luminosity while for $M_{\rm
BH}=6M_{\odot}$, the neutrino luminosity is comparable to or slightly
smaller than the electromagnetic luminosity.}

\begin{document}
\maketitle

\section{Introduction}

The latest numerical simulations show that the merger of black
hole-neutron star binaries and binary neutron stars often produces a
system composed of a black hole surrounded by a massive torus of mass
$\agt 0.1M_{\odot}$~\cite{BHNS,NSNS,LR,hotoke,SKKS11,REV}. Such
systems are promising candidates for the central engine of short-hard
gamma-ray bursts (SGRB)~\cite{SGRB0} which emit a huge amount of total
radiation energy $\sim 10^{48}$--$10^{50}$~ergs in a short time scale
$\alt 2$ s~\cite{SGRB}. The proposed mechanism for emitting the huge
radiation energy is the pair annihilation of neutrinos and
antineutrinos which are emitted from the hot and massive torus
surrounding the central black hole (e.g., Ref.~\citen{SJR}) or
electromagnetic mechanisms such as Blandford-Znajek mechanism, by
which rotational energy of a black hole can be efficiently
extracted~\cite{BZ,MG}, and/or Blandford-Payne mechanism (or similar
mechanisms) by which rotational energy of the torus can be extracted
by the magneto-centrifugal effect~\cite{BP,meier}.

To clarify the hypotheses that the black hole-torus system could be
the central engine of SGRB, a general-relativistic magnetohydrodynamic
(GRMHD) simulation including neutrino radiation effects is probably
the unique approach.  If we focus on the neutrino-pair-annihilation
scenario for generating the huge gamma-ray luminosity, the effect of the
neutrino radiation transport has to be taken into account. In a
previous paper~\cite{SST07}, we performed a GRMHD simulation (in a
fixed Kerr spacetime background) including microphysics effects as
well as neutrino cooling, extending the earlier Newtonian and
non-magnetohydrodynamics works (e.g., Refs.~\citen{SJR,LRP}).
However, to date, no simulation has been done in the framework of
general-relativistic, radiation-magnetohydrodynamics (GRRMHD).

In this paper, we report results of our first GRRMHD simulation for
the evolution of black hole-torus systems as a step toward a more
physical simulation incorporating with the detailed microphysics.  The simulation is
performed in full general relativity and in the framework of ideal
magnetohydrodynamics (MHD): We solve Einstein's equation, MHD
equation, induction equation, and a radiation transport equation.  
For evolving the radiation field of neutrinos, we employ a truncated moment
formalism as suggested in Refs.~\citen{AS,Kip,SKSS11}. For simplicity,
in this work, we do not consider the distribution in the frequency
space (spectrum) nor the flavor of neutrinos. We employ a
phenomenological equation of state (EOS) for the high-density (with
maximum density $\rho_{\rm max}=0.5$--$3.0 \times 10^{12}~{\rm
g/cm^3}$) torus matter and treat cross sections between matter and
neutrinos in an approximate manner. We believe that even with these
approximate treatments, this work {\em qualitatively} and {\em semi-quantitatively}
captures a
transport effect of neutrinos and coupling effects among general
relativity, MHD, and neutrino radiation transport for the first time.

This paper is organized as follows. In \S 2, the formulation for our
present GRRMHD simulation is summarized. In \S 3, the initial
condition for the black hole-torus system is given. In \S 4, EOS for
the matter in the torus, and coupling terms between the matter and
radiation are specified. We also describe our numerical method for the
evolution of the system. After a brief summary of quantities for
diagnostics in \S 5, numerical results are shown in \S 6, focusing in
particular on the luminosities of neutrinos and electromagnetic
radiations. Section 7 is devoted to a summary. In Appendixes A and B,
we present numerical results for test simulations to show that our
radiation hydrodynamic code is reliable.  Throughout this paper, the
geometrical units of $c=1=G$ are employed unless otherwise stated,
where $c$ is the speed of light and $G$ the gravitational constant.
Greek ($\alpha$, $\beta$, $\gamma \cdots$) and Latin ($i$, $j$, $k
\cdots$) subscripts denote the spacetime and space components,
respectively. The radiation quantities will be denoted by the Calligraphy.

\section{Formulation and basic equations}

\subsection{Framework}

We solve Einstein's equation, continuity equation for baryons, MHD equation,
induction equation, and a radiation transport equation all together
assuming the axial symmetry and the equatorial plane symmetry for the system
for the system.  
The basic equations are 
\beqn
&& G_{\mu\nu}=8\pi (T_{\mu\nu}^{\rm MHD}+{\cal T}_{\mu\nu}^{\rm rad}),\\
&& \nabla_{\mu}(\rho u^{\mu})=0,\\
&& \nabla^{\mu} T_{\mu\nu}^{\rm MHD}=- \cal{S}^{\rm rad}_{\nu},\\
&& \nabla_{\mu} F^{*\mu\nu}=0, \\
&& \nabla^{\mu} \cal{T}_{\mu\nu}^{\rm rad}=\cal{S}^{\rm rad}_{\nu},
\eeqn
where $G_{\mu\nu}$ is Einstein's tensor, $\rho$ is the rest-mass
density of baryon, $u^{\mu}$ is the four velocity, $\nabla_{\mu}$ is the
covariant derivative with respect to the spacetime metric
$g_{\mu\nu}$, $T_{\mu\nu}^{\rm MHD}$ and ${\cal T}_{\mu\nu}^{\rm rad}$ are
the stress-energy tensors of MHD and radiation
(neutrinos) parts, ${\cal S}^{\rm rad}_{\nu}$ is the source term of the radiation
field, and $F^{*\mu\nu}$ is the dual of the electromagnetic
tensor, respectively. In this paper, we assume that the ideal MHD
condition holds and write $F^{*\mu\nu}$ as
\beqn
F^{*\mu\nu}=b^{\mu} u^{\nu} - b^{\nu} u^{\mu},
\eeqn
where $b^{\mu}$ denotes the magnetic field vector defined in the fluid
rest frame, i.e., $b_{\mu}u^{\mu}=0$. 

The stress-energy tensors are written as
\beqn
&& T_{\mu\nu}^{\rm MHD}=(\rho h + b^2) u_{\mu} u_{\nu}
+\Big(P + {b^2 \over 2} \Big)g_{\mu\nu} - b_{\mu} b_{\nu},\\
&& {\cal T}_{\mu\nu}^{\rm rad}={\cal J} u_{\mu}u_{\nu} + {\cal H}_{\mu} u_{\mu}
+{\cal H}_{\nu} u_{\mu} + {\cal L}_{\mu\nu},\label{eq10}
\eeqn
where $h$ is the specific enthalpy defined by $1+\varep+P/\rho$ with
$\varep$ and $P$ being the specific internal energy and pressure, and
$b^2=b_{\mu}b^{\mu}$. ${\cal J}$, ${\cal H}_{\mu}$, and ${\cal L}_{\mu\nu}$ are the energy
density, momentum density, and stress tensor of the radiation field
\footnote{Our choice of physical symbols of the radiation field is 
different from those in a standard textbook such as Ref.~\citen{Miha}}, which 
are defined from the distribution function $f$ by \cite{SKSS11}
\beqn
&&{\cal J}=h_{\rm pl}^{4}
\int d\omega \omega^3 
\int f(\omega,\Omega,x^{\mu}) d\Omega,\\
&&{\cal H}^{\alpha}= h_{\rm pl}^{4}
\int d\omega \omega^3 
\int \ell^{\alpha}f(\omega,\Omega,x^{\mu}) d\Omega,\\ 
&&{\cal L}^{\alpha\beta}= h_{\rm pl}^{4}
\int d\omega\omega^3
\int \ell^{\alpha}\ell^{\beta} f(\omega,\Omega,x^{\mu}) d\Omega,
\eeqn
where $h_{\rm pl}$ is the Planck constant, $\omega$ is the frequency
of the radiation, $\int d\Omega$ denotes the integration over the
solid angle on unit sphere, and $\ell^{\alpha}$ is a spacelike unit
normal vector orthonormal to $u^{\alpha}$; $\ell^{\mu}\ell_{\mu}=1$
and $\ell_{\mu}u^{\mu}=0$. 
Note that we do not consider the evolution
of the spectrum for the radiation field for simplicity in this work. 
We also define the stress-energy tensor for the electromagnetic field by
\beqn
T_{\mu\nu}^{\rm EM}=b^2 u_{\mu} u_{\nu}
+{b^2 \over 2}g_{\mu\nu} - b_{\mu} b_{\nu}. 
\eeqn

\subsection{Einstein's equation}

Einstein's evolution equations are solved using the original version
of Baumgarte-Shapiro-Shibata-Nakamura formulation~\cite{BSSN} together
with the so-called moving-puncture
variable~\cite{puncture}. Specifically, we evolve a conformal factor
$W \equiv \gamma^{-1/6}$, the conformal three-metric
$\tilde{\gamma}_{ij} \equiv \gamma^{-1/3} \gamma_{ij}$, the trace of the
extrinsic curvature $K$, the conformal trace-free part of the
extrinsic curvature $\tilde{A}_{ij} \equiv \gamma^{-1/3} (K_{ij} - K
\gamma_{ij} / 3)$, and an auxiliary variable $F_i \equiv \tilde
\gamma_{ij,j}$. Here, $\gamma_{ij}$ and $K_{ij}$ are the spatial metric
and the extrinsic curvature, respectively.
For the evolution of the geometrical variables,
we implicitly assume to use the Cartesian coordinates;
the cartoon method is used for solving Einstein's equation\cite{cartoon}.

The spatial derivatives in the evolution equations are evaluated by a
fourth-order centered finite difference except for the advection terms
which is evaluated by a fourth-order up-wind finite difference. A
fourth-order Runge-Kutta method is employed for the time integration.
We employ a moving-puncture gauge for the lapse function $\alpha$ and
shift vector $\beta^k$ (e.g., Ref.~\citen{BB4}) in the form
\begin{eqnarray}
\partial_t \alpha  & = & - 2 \alpha K , \label{alpha} \\
\partial_t \beta^i & = & {3 \over 4} \tilde \gamma^{ij}
(F_j + \Delta t \partial_{t} F_j), \label{beta}
\end{eqnarray}
where $\Delta t$ is the time step.
% and $\dot F_j=\pa_t F_j$.  
We omit the advection terms in Eqs.~(\ref{alpha}) and (\ref{beta}) 
because black holes are located at the origin in the present setting 
and thus the advection terms do not play an important role. 

\subsection{MHD equations}

MHD equations are solved in the same numerical scheme as shown in
Ref.~\citen{SS05} to which the reader may refer. Specifically, we
choose the following variables to be evolved, 
\beqn
%% \rho_* = \rho w W^{-3},\\
&&S_0= W^{-3} T_{\mu\nu}^{\rm MHD} n^{\mu} n^{\nu},\\
&&S_i= - W^{-3} T_{\mu\nu}^{\rm MHD} n^{\mu} \gamma^{\nu}_{~i},
\eeqn
where $n^{\alpha}$ is the unit normal to spacelike hypersurfaces 
for which the components are
\beq
n^{\alpha}=\Big({1\over \alpha},-{\beta^k \over \alpha}\Big). 
\eeq
The Euler and energy equations are written in the forms
\beqn
&& \pa_t S_i + {1\over \sqrt{\eta}}
\pa_j (\sqrt{\eta} S_{ik} \gamma^{jk})
=-S_0 \pa_i \alpha +S_j \pa_i \beta^j - {1 \over 2}
\alpha S_{jk}\pa_i \gamma^{jk}
-\alpha W^{-3} {\cal S}^{\rm rad}_i,\\
&& \pa_t S_0 + {1\over \sqrt{\eta}}
\pa_j (\sqrt{\eta}\alpha S^j)=\alpha  K^{ij} S_{ij}
-S^i \pa_i \alpha + \alpha W^{-3} {\cal S}^{\rm rad}_{\alpha} n^{\alpha},
\eeqn
where $S_{ij}=W^{-3} T^{\rm MHD}_{\alpha\beta}\gamma^{\alpha}_{~i}
\gamma^{\beta}_{~j}$, $S^i=\gamma^{ij} S_j$, and $\eta$ is the 
determinant of the flat metric in the curvilinear coordinates.

The induction equation is solved defining a magnetic-field 
variable
\beqn
B^{\mu}= W^{-3}(w b^{\mu}-\alpha b^t u^{\mu}), 
\eeqn
where $w=\alpha u^t$ and $ B^{\mu} n_{\mu}=0$. We only need to
solve the spatial component $ B^i$ by 
\beqn
\pa_t  B^i={1\over \sqrt{\eta}}\pa_j [\sqrt{\eta}
( B^j v^i-  B^j v^i)],
\eeqn
where $v^i=u^i/u^t$ is the three velocity. 

The transport terms are specifically evaluated using a Kurganov-Tadmor
scheme~\cite{KT} with a piecewise parabolic reconstruction for the
quantities of cell interfaces.  The fourth-order Runge-Kutta method is
employed for the time integration.  For solving the induction
equation, a constrained-transport scheme with second-order
interpolation in space is employed~\cite{CT}.

\subsection{Radiation}

In the numerical simulation, instead of using Eq.~(\ref{eq10}), 
we rewrite ${\cal T}_{\alpha\beta}^{\rm rad}$ as \cite{SKSS11}
\beq
{\cal T}_{\alpha\beta}^{\rm rad}={\cal E}_{0} n_{\alpha} n_{\beta} 
+ ({\cal F}_{0})_{\alpha} n_{\beta} + ({\cal F}_{0})_{\beta} n_{\alpha}
+({\cal P}_{0})_{\alpha\beta}, \label{eq10a}
\eeq
and solve the equations for ${\cal E}_{0}$ and $({\cal F}_{0})_k$.  Here, 
$({\cal P}_{0})^{\alpha\beta}n_{\alpha}=0=({\cal F}_{0})^{\alpha}n_{\alpha}$.  
${\cal J}$ and ${\cal H}_{\alpha}$ are determined by
\beqn
&& {\cal J}={\cal E}_{0} w^2 - 2 ({\cal F}_{0})^{k} w u_{k} 
+ ({\cal P}_{0})^{ij} u_{i} u_{j},\\
&& {\cal H}^{\alpha}=({\cal E}_{0} w -({\cal F}_{0})^k u_k) h^{\alpha}_{~\beta} n^{\beta}
+w h^{\alpha}_{~\beta} ({\cal F}_{0})^{\beta}
-h^{\alpha}_{~i} u_j ({\cal P}_{0})^{ij}, \label{eq3.34}
\eeqn
%%%%%%%%
where $h_{\alpha\beta}=g_{\alpha\beta} + u_{\alpha} u_{\beta}$. We
here note $h^{\alpha}_{~\beta} n^{\beta}=n^{\alpha}-w u^{\alpha}$ and
$n_{\alpha}h^{\alpha\beta} \gamma_{\beta k} =-w u_k$. 
The introduction of ${\cal E}_{0}$ and $({\cal F}_{0})_i$ allows us to write 
the basic equations in a conservative form as
\beqn
&& \pa_t  {\cal E}
+{1\over \sqrt{\eta}}\pa_j[\sqrt{\eta} (\alpha {\cal F}^{j} -\beta^j
{\cal E})] \nonumber \\
&&\hskip 2cm 
=\alpha[{\cal P}^{ij}K_{ij}-{\cal F}^{j} \pa_j \ln \alpha 
- W^{-3} {\cal S}^{\rm rad}_{\alpha} n^{\alpha}],\label{eq1.3a}\\
%%%%%%%%%%%%%%%%%%%%%%%%%%%%%%%%%%%%
&&\pa_t {\cal F}_{i}
+{1\over \sqrt{\eta}}\pa_j[\sqrt{\eta} 
(\alpha {\cal P}_{i}^{~j} -\beta^j {\cal F}_{i})]
\nonumber \\
&&\hskip 2cm =\Big[-{\cal E} \pa_i \alpha + {\cal F}_{k}\pa_i \beta^k
-{\alpha \over 2}{\cal P}_{jk}\pa_i \gamma^{jk} + \alpha W^{-3} 
{\cal S}^{\rm rad}_i \Big],
\label{eq1.3b}
\eeqn
where ${\cal E}=W^{-3} {\cal E}_{0}$, ${\cal F}_i=W^{-3} ({\cal F}_{0})_i$, and 
${\cal P}_{ij}=W^{-3} ({\cal P}_{0})_{ij}$. Hence, the equations can be integrated in the
same numerical scheme (high-resolution shock capturing scheme) as the 
MHD equations. 

In the truncated moment formalism employed here, we have to impose a
closure relation for determining ${\cal P}_{ij}$. We
follow Ref.~\citen{SKSS11} on the prescription for this (we employ the
simplest version of Ref.~\citen{SKSS11}.)  For the optically thin
limit, we set
\beqn
({\cal P}^{ij})_{\rm thin}={\cal E} {{\cal F}^{i} {\cal F}^{j} \over
\gamma_{kl}{\cal F}^{k} {\cal F}^{l}}, \label{clos1}
\eeqn
and for the optically thick limit, 
\beqn
{\cal L}^{\alpha\beta}={1 \over 3} h^{\alpha\beta} {\cal J}. \label{clos2}
\eeqn
Equation~(\ref{clos2}) is rewritten in terms of ${\cal P}^{ij}$, 
${\cal E}$, and ${\cal F}_i$ as 
\beqn
({\cal P}^{ij})_{\rm thick} &=&
{{\cal E} \over 2w^2+1}\Big[(2w^2-1)\gamma^{ij}-4V^i V^j \Big]
+{1 \over w}({\cal F}^{i} V^j + {\cal F}^{j} V^i) \nonumber \\
&&~+{2{\cal F}^{k} u_k \over (2w^2+1)w}(-w^2 \gamma^{ij}+V^i V^j),
\eeqn
where $V^k=\gamma^{kl} u_l$. 

For the optically grey region, we assume 
\beqn
{\cal P}^{ij}={3\chi - 1\over 2}({\cal P}^{ij})_{\rm thin}
+{3(1-\chi) \over 2}({\cal P}^{ij})_{\rm thick}, \label{eqpij}
\eeqn
where $\chi$ is the so-called variable Eddington factor, which is 
$\chi = 1/3$ in the optically thick limit and $\chi=1$ in the 
optically thin limit. Following Ref.~\citen{Livermore}, we write 
$\chi$ as a function of $\hat {\cal F}$, for which we choose~\cite{SKSS11}
\beqn
{\hat {\cal F}}:=\biggl( {h_{\alpha\beta} 
{\cal H}^{\alpha}  {\cal H}^{\beta} 
\over {\cal J}^2}\biggr)^{1/2}. \label{edd2}
\eeqn
Following Livermore~\cite{Livermore}, we employ the following 
function for $\chi({\hat {\cal F}})$ as
\beqn
\chi={3+4{\hat {\cal F}}^2 \over 5+2\sqrt{4-3{\hat {\cal F}}^2}}. \label{chi1}
\eeqn

With the choice of (\ref{edd2}), $\hat F$ obeys an algebraic equation
for a given set of ${\cal E}$ and ${\cal F}_{j}$. This can be written
in the form
\beqn
\hat {\cal F}^2 = {h^{\alpha\gamma}{\cal T}_{\alpha\beta}^{\rm rad} u^{\beta}
{\cal T}_{\gamma\lambda}^{\rm rad}u^{\lambda} \over 
{\cal T}_{\alpha\beta}^{\rm rad} u^{\alpha}u^{\beta}},
\eeqn
where for ${\cal T}_{~\alpha\beta}^{\rm rad}$, Eq.~(\ref{eq10a}) is used with
Eq.~(\ref{eqpij}). In the numerical simulation, we solve this
algebraic equation numerically. 

%\subsection{Choice of the system, equation of state, and radiation source}

\section{Initial condition}

In this work, we perform simulations for the system composed of a
stellar-mass black hole surrounded by a high-density torus.  Such a
system is a possible outcome after the merger of binary neutron stars
and black hole-neutron star binaries. The latest simulations have
indeed shown that a massive disk may be formed after the merger (see,
e.g., Refs.~\citen{REV} for a review): For a total mass of the system
3--$8M_{\odot}$, the disk mass could be 0.1--$0.5M_{\odot}$ with the
maximum density $10^{11}$--$10^{13}~{\rm g/cm^3}$ (for binary
parameters and EOS which are favorable for a disk formation).
Generally speaking, the disk formation is enhanced if (i) the black
hole in the black hole-neutron star binary is rapidly
spinning~\cite{REV} or (ii) the EOS of neutron stars is stiff
(the radius of neutron stars is large) enough for the merger
to result in a long-lived hypermassive neutron star for the binary
neutron stars~\cite{hotoke,SKKS11}.

In this work, we set the black hole mass to be $M_{\rm BH}=3M_{\odot}$
or $6M_{\odot}$, and baryon rest-mass of the disk 
to be $M_{\rm disk}\approx0.14$--$0.38M_{\odot}$.  
The system with $M_{\rm BH}=3M_{\odot}$ is a model for the remnant of 
the binary neutron star merger and that with $M_{\rm BH}=6M_{\odot}$ is a 
model for black hole-neutron star binaries.  The nondimensional black 
hole spin is chosen to be $a=0$ for simplicity: In a previous paper~\cite{SST07}, we show that the
prograde spin enhances the neutrino luminosity, and hence, the results
of the present paper are likely to give a conservative estimate for
the neutrino luminosity. 
%%%%%%%%%%%For one model, we set $a=0.5$. %%%%%%
The maximum density $\rho_{\rm max}$ is chosen in the range
$2.3$--$3.1 \times 10^{12}~{\rm g/cm^3}$ for $M_{\rm BH}=3M_{\odot}$
and $5.1$--$7.5 \times 10^{11}~{\rm g/cm^3}$ for $M_{\rm
BH}=6M_{\odot}$.  
We note that for the same mass and radius of the disk in units of $M_{\rm BH}$,
the density is lower for the higher-mass black hole.
For these initial conditions, we prepare an
equilibrium state which is constructed by a code reported in
Ref.~\citen{S07} with the so-called $j$-constant law in which all the
fluid elements have the same specific angular momentum (the same value
of $h u_{\varphi}$).  Table~\ref{table1} lists key quantities for the
initial conditions.

\begin{table}[t]
 \caption{Key quantities for the initial conditions employed in the
numerical simulation. Black hole mass ($M_{\rm BH}$), nondimensional
spin parameter ($a$), location (in terms of the coordinate radius) of 
the inner and outer edges on the equatorial plane, baryon rest mass 
and maximum density of the torus,
maximum magnetic field strength, and maximum ratio of the magnetic
pressure to the gas pressure. $M_{\rm BH}$ and $M_{\rm disk}$ are 
shown in units of $M_{\odot}$. $r_1$ and $r_2$ are shown in 
units of $M_{\rm BH} (GM_{\rm BH}/c^{2})$.  
%For all the models, the black hole mass is $M_{\rm BH}=6M_{\odot}$. 
\label{table1}
}
%%%%%%%%%%%%%%
\begin{center}
\begin{tabular}{cccccccc} \hline
Model & $M_{\rm BH}$ 
& $a$ & $(r_1, r_2)$ & $M_{\rm disk}$
& $\rho_{\rm max} [{\rm g/cm^3}]$ & $B_{\rm max}$ (G)  
& $(P_B/P_{\rm gas})_{\rm max}$ \\
 \hline \hline
%%6a0m4l &6 & $0.0$ & $(4.07, 20.04)$ & $0.242$ & $3.64\times 10^{11}$ &
%%$3.5 \times 10^{14}$ & 1.5\% \\ \hline
6a0m4  &6 & $0.0$ & $(6.00, 17.65)$ & $0.242$ & $5.12\times 10^{11}$ &
$3.4 \times 10^{14}$ & 1.6\% \\ \hline
6a0m6  &6 & $0.0$ & $(6.00, 18.05)$ & $0.381$ & $7.45\times 10^{11}$ &
$3.6 \times 10^{14}$ & 1.5\% \\ \hline
%%6a5m5l &6 & $0.5$ & $(5.01, 20.01)$ & $0.297$ & $4.11\times 10^{11}$ &
%%$2.9 \times 10^{14}$ & 0.8\% \\ \hline
%%%%%%%%%%%%%%%%%%%%%%%%%%%%%%%%%%%%%%%%%%%%%
%%6a5m5  &6 & $0.5$ & $(6.00, 16.63)$ & $0.298$ & $7.06\times 10^{11}$ &
%%$4.6 \times 10^{14}$ & 1.6\% \\ \hline
3a0m1  &3 & $0.0$ & $(4.00, 18.05)$ & $0.145$ & $2.28\times 10^{12}$ &
$1.1 \times 10^{15}$ & 2.3\% \\ \hline
3a0m2  &3 & $0.0$ & $(4.00, 18.53)$ & $0.209$ & $3.08\times 10^{12}$ &
$1.0 \times 10^{15}$ & 1.6\% \\ \hline
\end{tabular}
\end{center}
\end{table}

At $t=0$, we superimpose a poloidal magnetic field inside the torus which
induces an angular momentum transport during the evolution as in, e.g.,
Refs.~\citen{MG,SST07}. Such an initial magnetic field configuration 
is chosen simply due to the technical reason: In the nearly vacuum region,
we cannot put magnetic fields for which the magnetic pressure is 
much larger than the (atmosphere) gas pressure because our MHD code is not allowed
to evolve an extremely low-$\beta$ plasma.
Note that even if the magnetic field profile is initially 
artificial, the resulting profile after several dynamical timescale 
evolution depends only weakly on the initial condition, as described 
in many papers in this field (e.g., Ref~\citen{MG}): 
The resulting magnetic fields are usually composed of a random magnetic 
field inside the torus and of approximately dipole magnetic fields outside the torus and black 
hole magnetosphere. These seem to be a universal outcome after the onset 
of magnetohydrodynamical instabilities. Thus we adopt a simple magnetic 
field profile initially.

The profile for the toroidal component of the vector potential of 
the magnetic field is chosen as
\beqn
A_{\varphi}=\left\{
\begin{array}{ll}
A_0(P-P_0) & {\rm for}~P > P_0,\\
0 & {\rm for}~P \leq P_0,
\end{array}
\right.
\eeqn
and then the magnetic field is given by
\beqn
B^i=W^{-3}\epsilon^{ijk} \pa_j A_k,
\eeqn
where $\epsilon^{ijk}$ is the completely antisymmetric tensor.  $P_0$
is a parameter chosen to be $0.04P_{\rm max}$ where $P_{\rm max}$ is
the maximum pressure. $A_0$ is the constant that determines the
magnetic field strength. In this paper, we give a strong magnetic
field initially to make the angular momentum transport turn on immediately
after the onset of simulations. The given maximum magnetic field
strength is $B_{\rm max} \approx 3$--$5 \times 10^{14}$~G for $M_{\rm
BH}=6M_{\odot}$ and $\approx 1\times 10^{15}$~G for $M_{\rm
BH}=3M_{\odot}$, with which the maximum ratio of the magnetic pressure
to the gas pressure is $\sim 1.5$--2.3\% (see Table~\ref{table1}). A
latest GRMHD simulation for the binary neutron star merger indicates
that such a strong magnetic field will be achieved during the
evolution of accretion torus surrounding central black holes
\cite{LR}.

We employed the different field strength for a few models as a test and found that the
results depend very weakly on it as far as the field strength is
sufficiently large. For a weak initial field strength, by contrast,
the angular momentum transport does not proceed efficiently, and
hence, the degree of the subsequent turbulent motion is weak and the
efficiency of the shock heating is also low. The likely reason is that
with the weak field strength, the fastest growing mode of the
magnetorotational instability (MRI) is not resolved. The strong
magnetic field has to be set up to enhance the efficient angular
momentum transport and shock heating in the present grid resolution.

\section{Equation of state and treatment of radiation}

\subsection{Equation of state}

For a high-density matter with $\rho \sim 10^{10}$--$10^{13}~{\rm
g/cm^3}$, the dominant pressure sources are degenerate electrons and
thermal nucleons for the typical temperature $10^{10}$--$10^{11}$~K.
The pressure for them is written, respectively, by
\beqn
&&P_{\rm deg}=4.93 \times 10^{30} 
\Big({Y_e \rho_{12} \over 0.5}\Big)^{4/3}~{\rm g/cm/s^{2}},\\
&&P_{\rm gas}=8.31 \times 10^{29} \rho_{12} T_{10}~{\rm g/cm/s^{2}},
\eeqn
where $\rho_{12}=\rho/(10^{12}~{\rm g/cm^3})$, $Y_e$ is the fraction of
electrons per nucleon, and $T_{10}=T/(10^{10}$~K).  
%%$P_{\rm deg}$ and $P_{\rm gas}$ denote the pressure due to degenerate
%%electrons and gas pressure due to free nucleon, respectively. 
We here assume that the baryon is composed only of free nucleons. The
{\it radiation} pressure due to photons and electron-positron pairs is
written as
\beqn
P_{\gamma}=\Big( {1 \over 3} + {7 \over 12} \Big)a_{\rm r}T^4
=7.04 \times 10^{25} T_{10}^4~{\rm g/cm/s^{2}}, \label{radpre}
\eeqn
where $a_{\rm r}$ is the radiation density constant ($7.56 \times
10^{-15}~{\rm ergs/cm^3/K^4}$).
The effect of the radiation pressure on dynamics is minor in the main body of 
the torus where $\rho_{12} \agt 1$, as far as $T_{10} \alt 10$, although
the radiation pressure plays a role for a high-temperature and relatively 
low-density region that does not contribute much to the neutrino luminosity. 
Hence, we ignore its contribution to the pressure.

Thus, for simplicity, we employ the following EOS:
\beqn
%%P=K \rho^{4/3} + {\rho \over m_u} k_{\rm B} T,
P=K \rho^{4/3} + (\Gamma_{\rm th}-1) \rho \varepsilon_{\rm th}, 
\eeqn
where $K$ is a polytropic constant which is $4.93 \times 10^{14}
(Y_e/0.5)^{4/3}$ (in the cgs unit) for degenerate relativistic electrons~\cite{ST},
and $\Gamma_{\rm th}$ is the adiabatic constant for which we choose
$5/3$, because free nucleons mainly contribute to the thermal part.
$\varep_{\rm th}$ is the specific thermal energy determined by
\beqn
\varep_{\rm th}=\varep - \varep_{\rm poly},
\eeqn
where $\varep_{\rm poly}=3K\rho^{1/3}$ is the specific internal energy
associated with degenerate electrons.  In this paper, we give a fixed
value for $Y_e$ as 0.3; we suppose a moderately neutron-rich material.

The matter temperature may be determined from the relation
\beq
k_{\rm B} T = {2\over 3} m_u \varep_{\rm th}, 
\eeq
where $m_u$ is the atomic mass unit ($m_u=1.66057 \times 10^{-24}$~g), 
and $k_{\rm B}$ is the Boltzmann constant. 
Specifically, \footnote{For clarifying the unit, we recover $c$ and
$G$ in this section.} 
\beqn
T=7.206 \times 10^{10}\Big({\varep_{\rm th} \over 0.01c^2}\Big)~{\rm
K}. \label{temper}
\eeqn
In reality, the temperature does not increase linearly with
$\varep_{\rm th}$, because the contribution of the radiation (photons
and electron-positron pairs) to the internal energy in fact 
becomes significant for high temperature; see Eq.~(\ref{radpre}). 
Instead of solving the 4th order nonlinear equation for $T$, 
we simply set a maximum value for the temperature to be 
$k_{\rm B}T_{\rm max}=30$~MeV: A latest fully general relativistic 
simulation shows that this is indeed a reasonable upper bound~\cite{SKKS11}. 
If the value of $T$, determined from Eq.~(\ref{temper}), exceeds 
$T_{\rm max}$, we set $T=T_{\rm max}$. 
This procedure may give a higher estimation of $T$ for 
$10 \alt T_{10} \alt 30$.
However, the density of such high temperature region is rather small
as $\rho \alt 10^{8-9}$ g/cm$^{3}$ (see the top and middle panels of Fig.~\ref{fig1}), 
and hence, spurious effects of this simple procedure seem to be minor.

\subsection{Radiation source term}

The source term for the radiation transport equation, $S^{\rm rad}_{\alpha}$, 
is determined in the following manner.  For neutrinos
in the medium with $\rho \sim 10^{11}$--$10^{13}~{\rm g/cm^3}$, the
dominant process is the absorption by
nucleons~\cite{TS74,Bruen,Rampp}.  Taking into account this fact, we
write it as
\beqn
{\cal S}^{\rm rad}_{\alpha}=\kappa [({\cal J}^{\rm eq}(T)-{\cal J})u_{\alpha} -{\cal H}_{\alpha}],
\eeqn
where we write for simplicity, 
\beqn
{\cal J}^{\rm eq}={7 \over 8} a_{\rm r} T^4, \label{eqjeq}
\eeqn
assuming that neutrinos are close to the $\beta$-equilibrium.  This
assumption is reasonable in the main part of the torus.  Note that $T$
in Eq.~(\ref{eqjeq}) is the matter temperature.  We here assume that
only electron neutrinos and anti-neutrinos are present.  Because we
omit other sources for the neutrino emission such as pair production
process and also emission of muon and tau neutrinos, the total
neutrino luminosity is likely to be underestimated. However, the
present treatment is acceptable for a qualitative and semi-quantitative study, e.g., to
derive an approximate order of magnitude of the neutrino luminosity.
$\kappa$ is the opacity and we here define by $\rho \sigma c /m_u$
where $\sigma$ is the cross section of neutrinos with matter. For
simplicity, we write it
\beqn
\sigma=C_{\sigma}\sigma_{\rm w} \Big({k_{\rm B} T \over m_e c^2}\Big)^2,
\eeqn
where $\sigma_{\rm w}=1.76 \times 10^{-44}~{\rm cm^2}$ denotes the
characteristic cross section of weak interactions~\cite{ST}, $m_e c^2$
is the electron rest-mass energy, 511~KeV, and $C_{\sigma}$ is a
constant which we choose 1; we choose $C_{\sigma}=0.5$ and 2 for test
simulations, but the results (evolution of torus and neutrino luminosity) 
depend only weakly on this parameter.
The dependence on the neutrino energy is simply estimated by the local
matter temperature $k_{\rm B}T$, because we do not have the method for
estimating it in the present treatment. In the optically thick region,
our prescription is acceptable.  However, in the grey and optically
thin regions, a more sophisticated treatment is required.

%$k_{\rm B}T$ here should be a typical neutrino energy, but we do not
%have the method for estimating it. For simplicity, we input the local
%matter temperature focusing on the region where the absorption plays
%an important role (i.e., the region where the optical depth is high).

For this choice of $\sigma$, a nondimensional quantity $\kappa GM_{\rm
BH}/c^3$ is written as
\beqn
{\kappa GM_{\rm BH} \over c^3}
=3.6 C_{\sigma}\rho_{12}\Big({k_{\rm B} T \over 10~{\rm
MeV}}\Big)^2 \Big({M_{\rm BH} \over 6M_{\odot}}\Big). 
\eeqn
Thus, a torus of thickness $\sim 10GM_{\rm BH}/c^2$ and of temperature
$k_{\rm B}T =10$~MeV surrounding a black hole of mass $6M_{\odot}$ is 
optically thick for $\rho \agt 3
\times 10^{10}~{\rm g/cm^3}$ with $C_{\sigma}=1$. 

\subsection{Partial implicit scheme}

The timescale of the neutrino absorption and emission $\approx 
\kappa^{-1}$ is often shorter than the dynamical timescale of the
system $\sim GM_{\rm BH}/c^3 \sim 0.03$~ms. This prohibits the
explicit time integration of the radiation transport equations. For a
stable numerical computation, we employ a partial implicit
(explicit-implicit) scheme for the time integration.  Schematically,
Eqs.~(\ref{eq1.3a}) and (\ref{eq1.3b}) are written in a matrix form 
\beqn
\pa_t \mbox{\boldmath$F$}=\mbox{\boldmath$T$} 
+ \kappa (\mbox{\boldmath$S$}_0 -\mbox{\boldmath$A$}\mbox{\boldmath$F$}),
\eeqn
where $\mbox{\boldmath$F$}$ denotes a vector composed of 
\beqn
\mbox{\boldmath$F$}=\left(
\begin{array}{l}
{\cal E}   \\
{\cal F}_x \\
{\cal F}_y \\
{\cal F}_z \\
\end{array}
\right).
\eeqn
$\mbox{\boldmath$T$}$ denotes the sum of the transport term $\pa_i
(\cdots)^i$ and the source term associated with the gravitational
fields.  $\mbox{\boldmath$S$}_0$ denotes the source term composed of
the thermal quantity $a_{\rm r}T^4$, and $\mbox{\boldmath$A$}$ is a $4
\times 4$ matrix composed of the hydrodynamic and geometric 
quantities.  In each Runge-Kutta time integration, the term $\pa_t
\mbox{\boldmath$F$}$ is discretized as $(\mbox{\boldmath$F$}^{n+1}-
\mbox{\boldmath$F$}^n)/\Delta t$ where $n$ and $n+1$ denote 
neighboring two time steps for the fourth-order Runge-Kutta time
integration ($n=0$--3).  In the partial implicit scheme employed here,
$n$-th quantities are assigned for $\mbox{\boldmath$T$}$,
$\mbox{\boldmath$S$}_0$, $\mbox{\boldmath$A$}$, and $\kappa$, while we
assign ($n+1$)-th quantities for $\mbox{\boldmath$F$}$ in the
right-hand side. Namely, we write the equation in the following form:
\beqn 
(1 + \kappa^n \Delta t \mbox{\boldmath$A$}^n)\mbox{\boldmath$F$}^{n+1} =\mbox{\boldmath$F$}^n +
\Delta t (\mbox{\boldmath$T$}^n + \kappa^n \mbox{\boldmath$S$}_0^n). 
\eeqn 
This is a simple $4 \times 4$ matrix equation and solved in a
straightforward manner.

After the radiation transport equations are integrated, we integrate
the MHD equations. The transport term and ordinary source term
associated with the gravitational field of this equation are handled
in the same manner as in our previous paper; to obtain the ($n+1$)-th
quantities, $n$-th quantities are assigned for these terms.  In the
right-hand side of the MHD equations, the source terms composed of the
radiation fields are included additionally in the present case;
$-\kappa (\mbox{\boldmath$S$}_0 -
\mbox{\boldmath$A$}\mbox{\boldmath$F$})$.  We handle this term as in
the radiation transport equation, i.e., in the following time
stepping:
\beq
-\kappa^n (\mbox{\boldmath$S$}_0^n - \mbox{\boldmath$A$}^n 
\mbox{\boldmath$F$}^{n+1}).
\eeq
This implies that the absolute value of the source term associated
with the radiation field in the MHD equations is equal to that in the
radiation transport equation, and hence, the energy and momentum
contributed from this source term cancel each other for the total
conservation equation $\nabla^{\mu} (T_{\mu\nu}^{\rm MHD}+{\cal T}_{\mu\nu}^{\rm rad})=0$. 

In Appendix A, we present results of a test simulation using a
semi-analytic solution of radiation-hydrodynamical Bondi flow.  In
Appendix B, we also present results of one-dimensional test
simulations of Riemann problems in special relativity.  We show that by
our code, it is feasible to stably perform a radiation hydrodynamics
simulation and also to derive a second-order convergent result for 
high values of $\kappa$ (for the optically thick case) for 
the general relativistic flows. 

\section{Quantities for diagnostics}

We monitor the total baryon rest-mass $M_*$, %angular momentum $J$,
internal energy $E_{\rm int}$, thermal component of internal energy
$E_{\rm th}$, kinetic energy $T_{\rm kin}$,
%and rotational kinetic energy $T_{\rm rot}$ for baryons as well as 
and electromagnetic energy $E_{\rm B}$ by
\begin{eqnarray}
&&M_*= \int \rho_* \sqrt{\eta} d^3x, \label{eq5.1}\\
%%&&J = \int \rho_* h u_{\varphi} \sqrt{\eta} d^3 x, \\
&&E_{\rm int} = \int \rho_* \varepsilon \sqrt{\eta} d^3x , \\
&&E_{\rm th} = \int \rho_* \varepsilon_{\rm th} \sqrt{\eta} d^3x , \\
&&T_{\rm kin} = \int {1\over 2} \rho_* h  v^i u_{i}
\sqrt{\eta} d^3x, \label{eq:Tkin} \\
%%%&&T_{\rm rot} = \int {1\over 2} \rho_* h \Omega u_{\varphi}
%%%\sqrt{\eta} d^3x, \label{eq:Trot} \\
&&E_{\rm B}  = \int T^{\rm EM}_{\mu\nu} n^{\mu}n^{\nu} \alpha^{-1} 
\sqrt{\eta} d^3x. \label{eq5.5}
\end{eqnarray}
%where $\Omega$ is the angular velocity defined by $u^{\varphi}/u^t$.
Here, for the definition of $E_{\rm int}$, $T_{\rm kin}$, and $E_{\rm B}$,
we follow Ref.~\citen{DLSSS2}. Note that $M_{*}=M_{\rm disk}$ at initial.

In stationary axisymmetric spacetimes, the following relations are
derived from the conservation law of the energy-momentum tensor in the
absence of neutrino cooling:
\beqn
\pa_{\mu} (\sqrt{-g} (T^{{\rm MHD}})^{\mu}_{~t})=0.\label{tmunut}
%\\
%\pa_{\mu} (\sqrt{-g} (T^{{\rm MHD}})^{\mu}_{~~\varphi})=0. \label{tmunup} 
\eeqn
From this relation, it is natural to define the ejection rate of 
energy 
\beqn
&&\dot E =-\oint_{r={\rm const}} (T^{{\rm MHD}})^{r}_{~t} \sqrt{-g}
dS,
%%\\ 
%%&&\dot J =\oint_{r={\rm const}} (T^{{\rm MHD}})^{r}_{~\varphi}
%%\sqrt{-g} dS,
\eeqn
where $dS=d\theta d\varphi$ and the surface integral is performed 
for a two surface far from the torus. 
From the continuity equation, the rest-mass ejection rate is defined in the 
same manner as
\beqn
&&\dot M_*=\oint_{r={\rm const}} \rho_* v^r r^2 dS. 
\eeqn

Emission rates of energy 
%%and angular momentum 
for the electromagnetic field and radiation are defined in the same
manner as
\beqn
&&\dot E_{\rm B} =-\oint_{r={\rm const}} (T^{{\rm EM}})^{r}_{~t}
\sqrt{-g} dS,\\ 
%&&\dot J_{\rm B} =\oint_{r={\rm const}} (T^{{\rm EM}})^{r}_{~\varphi}
%\sqrt{-g} dS, \\
&&\dot E_{\rm rad} =-\oint_{r={\rm const}} ({\cal T}^{{\rm rad}})^{r}_{~t}
\sqrt{-g} dS.
%&&\dot J_{\rm rad} =\oint_{r={\rm const}} (T^{{\rm
%rad}})^{r}_{~\varphi} \sqrt{-g} dS. 
\eeqn

Pair annihilation rates of neutrinos and antineutrinos are roughly
evaluated in the following manner: First we assume that the luminosity
of neutrinos and antineutrinos are $c_{\nu} e$ and $(1-c_{\nu}) e$,
respectively.  Here, $c_{\nu}$ is a constant for which $0 < c_{\nu} <
1$.  Then the pair annihilation luminosity per volume is approximately
estimated by~\cite{bel}
\footnote{We here recover $c$ for clarifying the physical units.}
\beqn
\dot q_{\nu\bar\nu}= c_{\rm max} c \sigma_0 {e^2 \over (m_e c^2)^2}
\sqrt{E_{\nu}E_{\bar\nu}},\label{eq5.13}
\eeqn
where $c_{\rm max}={\rm max}[c_{\nu}^2, (1-c_{\nu})^2]\cos^2\Theta$
with $\Theta$ being the collision angle of pairs. Note that the
maximum value of $c_{\rm max}$ is $1/4$.  The typical value of
$\Theta$ is determined by the configuration of the neutrino-emission
region: If the pair annihilation occurs in the vicinity of the black
hole and rotation axis, $\cos\Theta$ is likely to be near the
unity. $\sigma_0(\approx 3.3 \times 10^{-45}~{\rm cm}^2)$ is the
typical cross section of the pair annihilation, and $E_{\nu}$ and
$E_{\bar\nu}$ are the characteristic energy of neutrinos and
antineutrinos, respectively.  From $\dot q_{\nu\bar\nu}$, we define an
approximate pair annihilation luminosity by
\beqn
L_{\nu\bar\nu}=\int W^{-3}dV \dot q_{\nu\bar\nu}. 
\eeqn
In the following, we show the result of $L_{\nu\bar\nu}$ with $c_{\rm
max}\sqrt{E_{\nu}E_{\bar\nu}}/(m_e c^2)=1/4$ as a reference for the
order of magnitude of $L_{\nu\bar\nu}$.

\section{Numerical results}

\begin{figure}[t]
\begin{center}
\epsfxsize=2.2in
\leavevmode
\epsffile{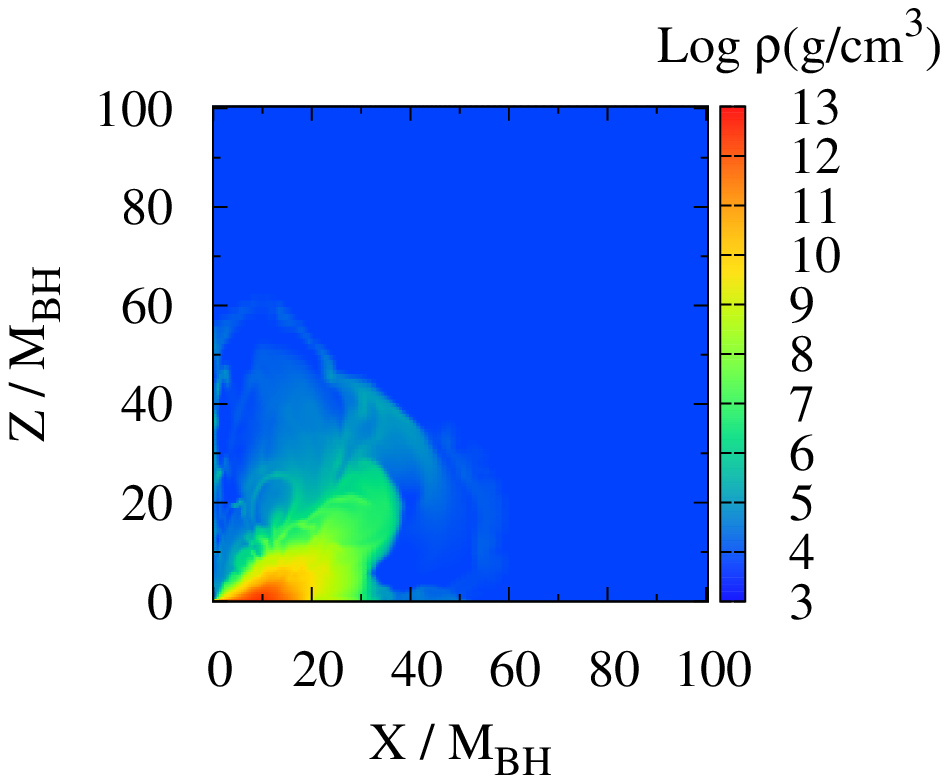}
\epsfxsize=2.2in
\leavevmode
\hspace{-1.7cm}
\epsffile{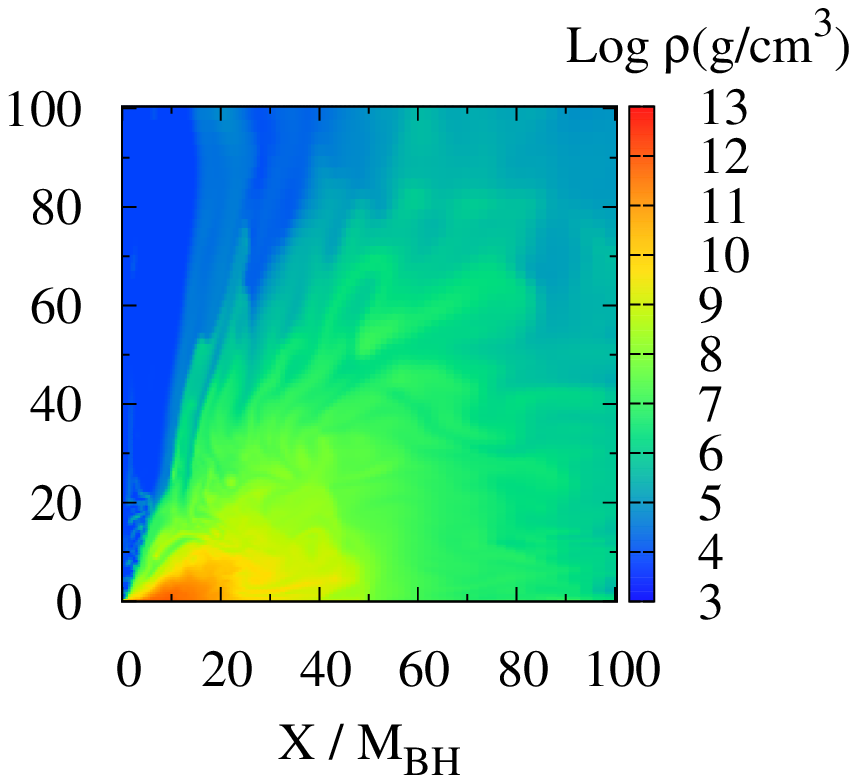}
\epsfxsize=2.2in
\leavevmode
\hspace{-1.7cm}
\epsffile{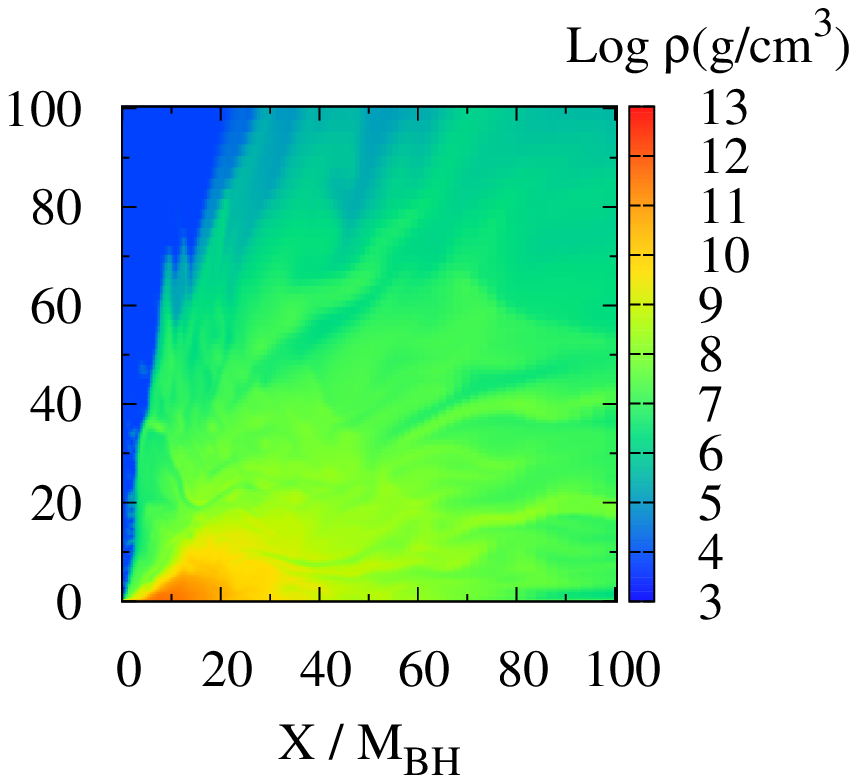}\\
\epsfxsize=2.2in
\leavevmode
\epsffile{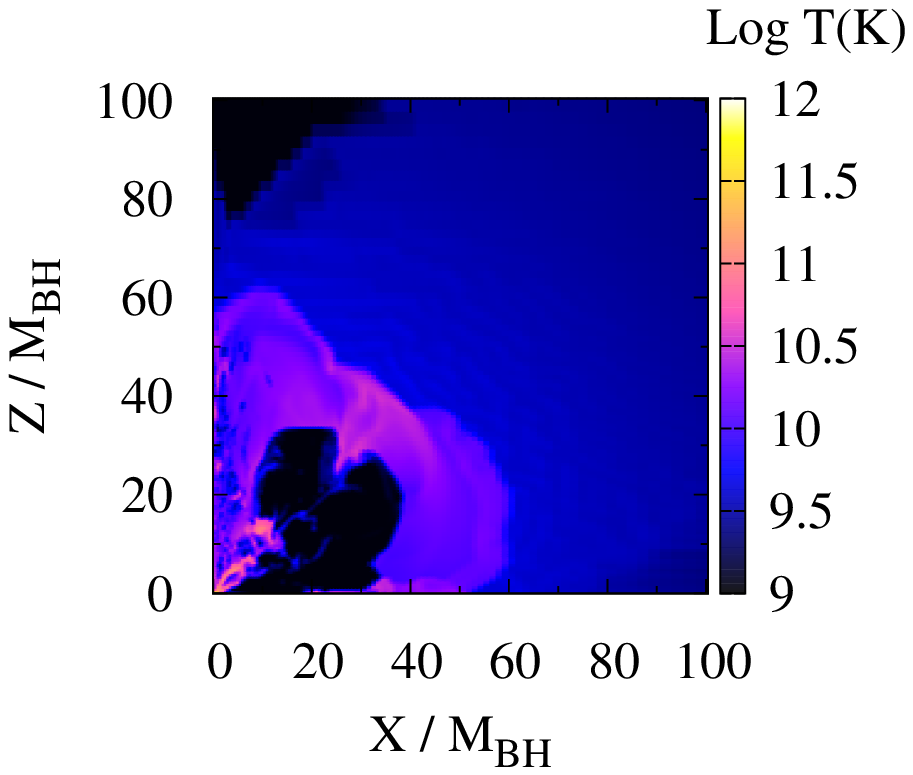}
\epsfxsize=2.2in
\leavevmode
\hspace{-1.7cm}
\epsffile{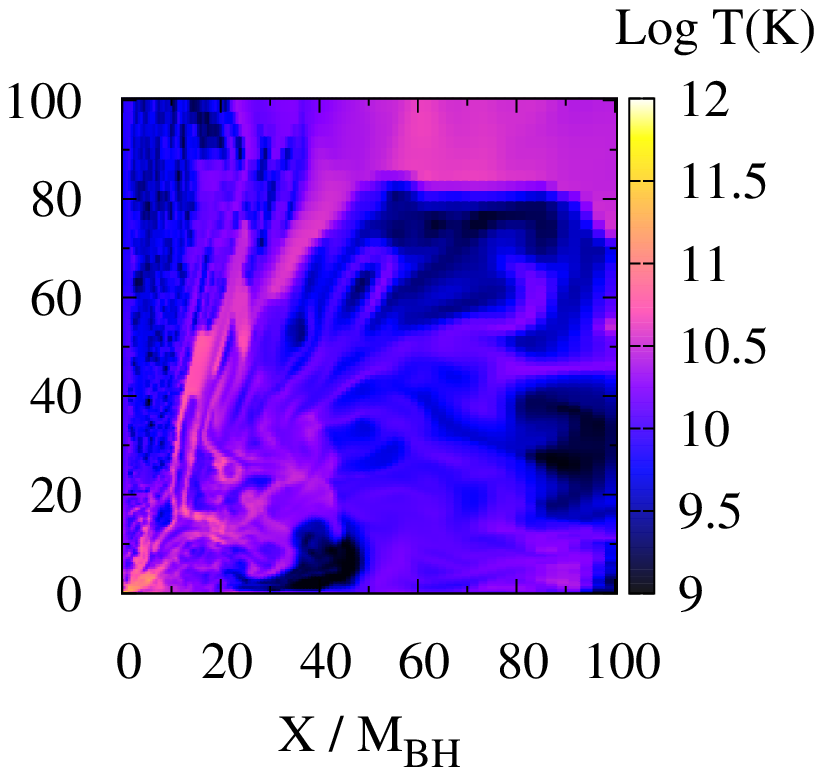}
\epsfxsize=2.2in
\leavevmode
\hspace{-1.7cm}
\epsffile{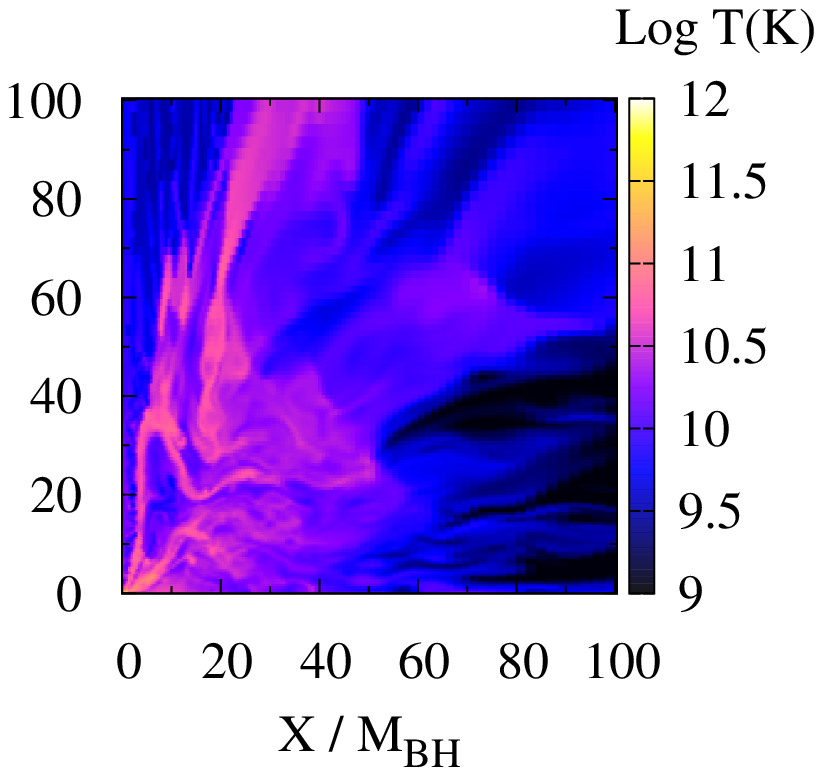} \\
\epsfxsize=2.2in
\leavevmode
\epsffile{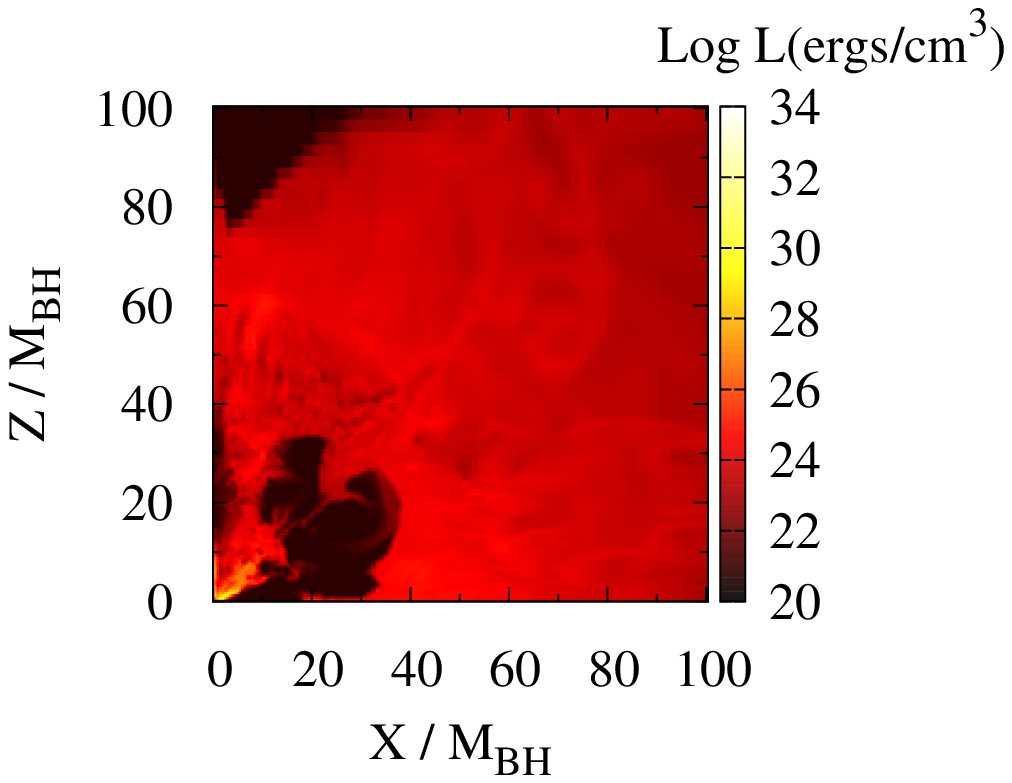}
\epsfxsize=2.2in
\leavevmode
\hspace{-1.7cm}
\epsffile{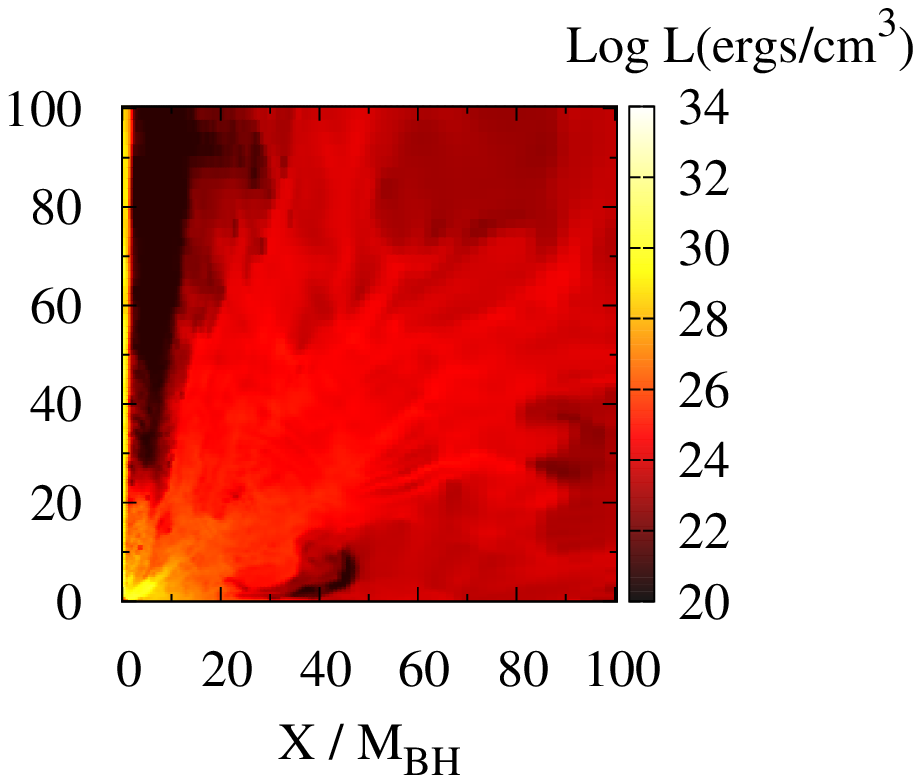}
\epsfxsize=2.2in
\leavevmode
\hspace{-1.7cm}
\epsffile{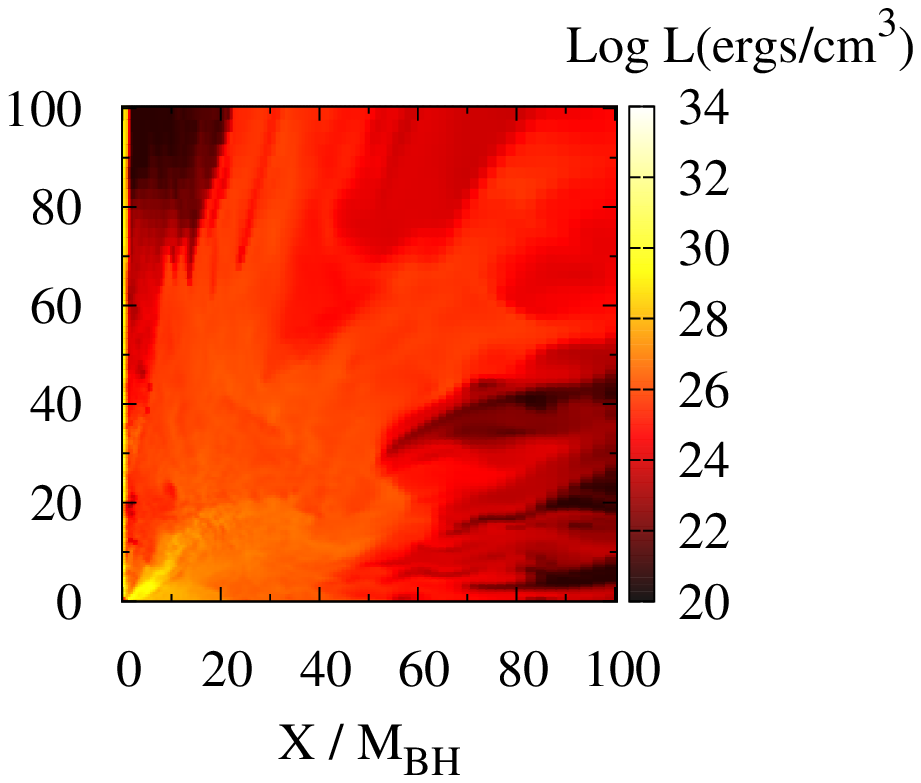}
\vspace{-3mm}
\caption{
Snapshots for the distribution of the rest-mass density (top panels),
matter temperature (middle panels), and radiation field density
(${\cal E}_{0}={\cal E}W^3$, bottom panels) at selected time slices, $t=5.7$~ms 
(left), 21.3~ms (middle), and 42.5~ms (right), for model 3a0m2. 
\label{fig1}}
\end{center}
\end{figure}

\begin{figure}[t]
\begin{center}
\epsfxsize=2.6in
\leavevmode
\epsffile{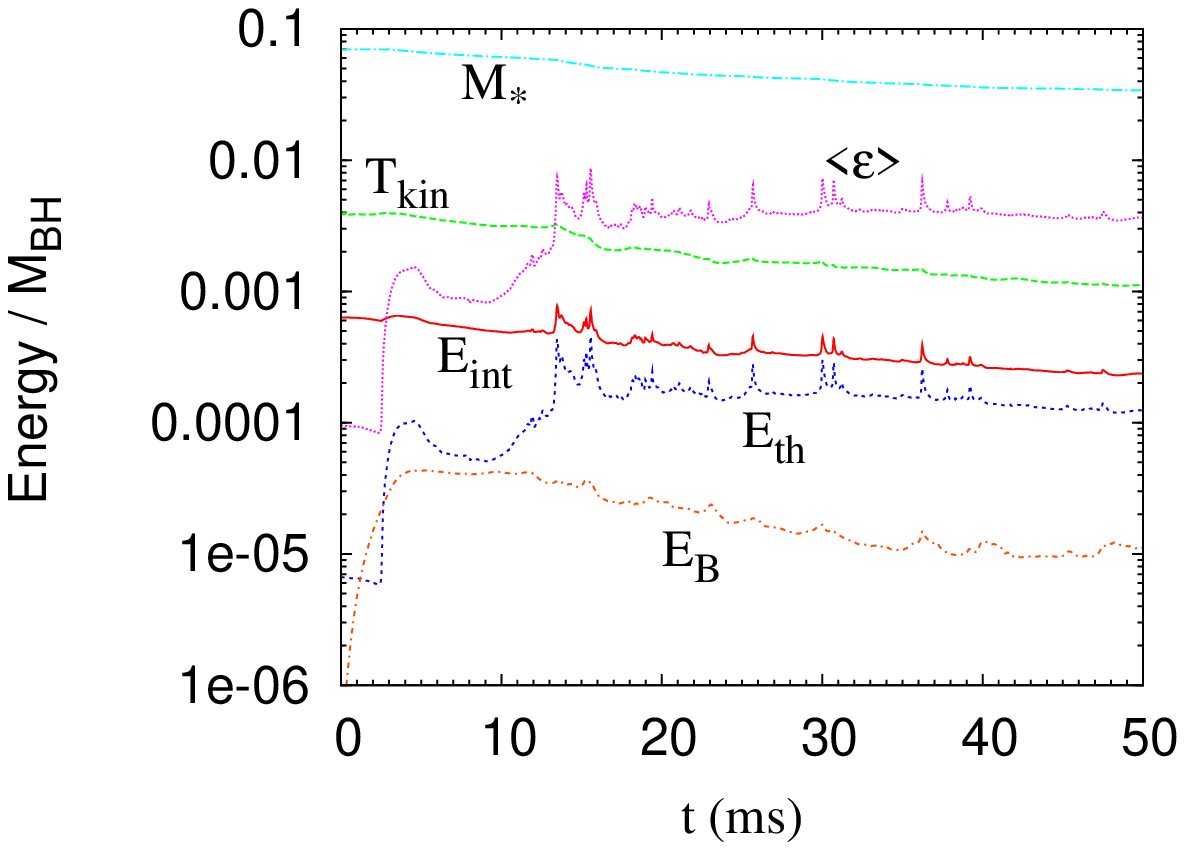}
\epsfxsize=2.6in
\leavevmode
\epsffile{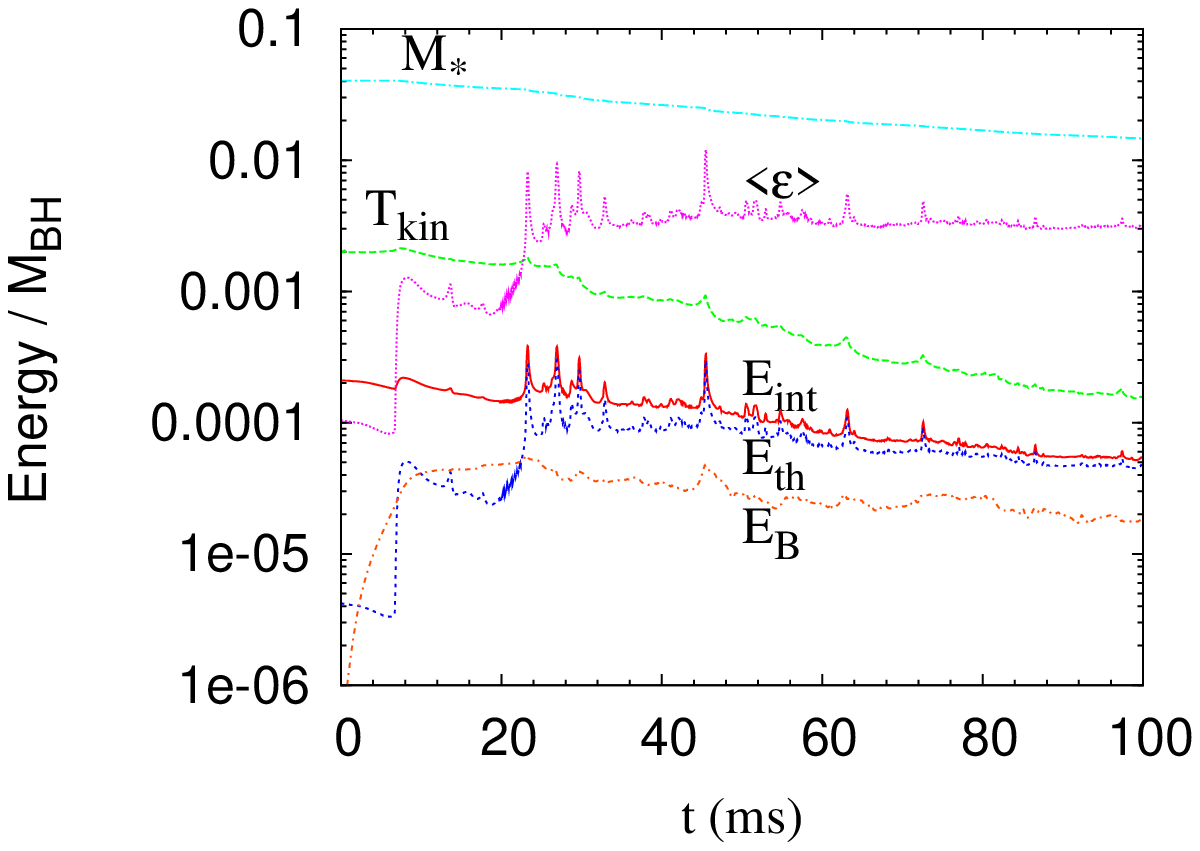}
\vspace{-3mm}
\caption{
Evolution of the total rest-mass energy ($M_*$), kinetic energy
($T_{\rm kin}$), internal energy ($E_{\rm int}$), thermal energy
($E_{\rm th}$), and electromagnetic energy ($E_{\rm B}$) measured
outside the apparent horizon for models 3a0m2 (left) and 6a0m4 (right)
as in Fig.~\ref{fig1}. We also plot an averaged value of the specific
internal energy defined by $\langle \varepsilon \rangle=E_{\rm
int}/M_*$.
\label{fig2}}
\end{center}
\end{figure}

\subsection{Setting}

Numerical simulations are performed in a nonuniform grid with
the $(x,z)$ coordinates. In addition, we prepare 5 grid points along the
$y$ direction for solving Einstein's equation by the Cartoon
method~\cite{cartoon} with the fourth-order accuracy.  The grid
spacing for each coordinate at $i$-th grid point is determined by the
following rule,
\beqn
\Delta x_i=\left\{
\begin{array}{ll}
\Delta x_0 & {\rm for}~0 \leq x \leq x_{\rm in},\\ 
C_x \Delta x_{i-1} & {\rm for}~x \geq x_{\rm in},\\ 
\end{array}
\right.
\eeqn
where $\Delta x_{0}$ is a constant, $\Delta x_i=x_{i+1}-x_i$, 
and $x_i$ is the $i$-th grid point with $0 \leq i \leq N$. 
Here, $x_i$ denotes $x$ or $z$. 
For $y$, the grid is set at $\pm \Delta x_{0}$ and $\pm 2\Delta x_{0}$.
$C_x$ is a constant which
is set to be 1.016. We employ $(N, \Delta x_0)=(630, 0.05 M_{\rm BH})$
and $(500, 0.075 M_{\rm BH})$ for the high and low grid resolutions,
respectively. $x_{\rm in}=20M_{\rm BH}$ and $22.5M_{\rm BH}$ for
$N=630$ and 500. With this setting, the outer boundary along each axis
for the high- and low-resolution runs is located at $x_N=139M_{\rm
BH}$ and $132M_{\rm BH}$, respectively. By several test simulations,
we confirmed that the numerical results depend only weakly on the grid
resolution.\footnote{We here imply that the results depend only weakly
on the grid resolution {\em in the time-averaged sense}.  Because the
matter and magnetic fields violently vary due to a turbulent motion
induced by magnetohydrodynamic instabilities, the results do not agree well with
each other for the comparison done at a given instance of time.}  Hence, we only show the
results for the higher-resolution runs in the following.

\subsection{General features of dynamics}

In the {\em axisymmetric} simulation, the magnetic field eventually
escapes from the torus after the amplification of the magnetic-field
strength saturates. The reason is that the dynamo action is not
sustained in the axisymmetric system because of the anti dynamo
property~\cite{moffatt78}. Thus, an extremely long-term simulation does not
provide a realistic result because the angular momentum transport
process induced by the MHD effect does not work in its late
phase. Keeping in mind this fact, we always stopped the simulations at
$t=3500M_{\rm BH}$ following Ref.~\citen{MG}.  This corresponds to
$\approx 50$~ms and 100~ms for $M_{\rm BH}=3M_{\odot}$ and
$6M_{\odot}$, respectively.

Figure~\ref{fig1} plots the snapshots for the distribution of the
rest-mass density, matter temperature, and radiation field density
(${\cal E}_{0}={\cal E}W^3$) at $t=5.7$, 21.3, and 42.5 ms for model 3a0m2. This
shows the universal qualitative features for the evolution of the
system: After the simulation is started, the magnetic-field strength
increases due to the winding and MRI caused by the differential
rotation (see Fig.~\ref{fig2}). These processes also play an active
role for transporting angular momentum from inner parts to outer parts
of the torus. Due to the increase of the magnetic-field pressure, the
matter is ejected from the torus predominantly toward the high
latitude (see the left panels of Fig.~\ref{fig1}). By this process,
the geometrical thickness of the matter distribution increases,
resulting in formation of a quasisteady funnel structure which is
composed of the geometrically thick torus and a nearly vacuum region
in the vicinity of the rotational axis (see the middle and right
panels of Fig.~\ref{fig1}). Due to the angular momentum transport
process, on the other hand, the matter for which angular momentum is
removed falls into the BH, and the total mass of the torus gradually
decreases.  A part of the matter ejected from the torus to a high
latitude comes back to the torus. If the angular momentum of the
matter is extracted by the magnetic processes during its travel, the
matter eventually comes back to an inner region of the torus.  When it
hits the torus, a shock is generated and its kinetic energy is
liberated. As a consequence of this process which continuously occurs,
the temperature of the inner region of the torus increases to a value
larger than 10~MeV, and such a region becomes a primary emission
region of neutrinos.

Because a funnel structure is formed and the neutrino emission is most
active near the inner region of the torus, neutrinos are emitted
primarily toward the outward direction in the vicinity of the rotational axis. The
bottom panels of Fig.~\ref{fig1} indeed shows that the neutrino radiation
density is high in a cone around the rotational axis and the opening
angle of the cone is rather small (cf. Fig.~\ref{fig3}).  Due to this collimation
property, the radiation density is enhanced.  This is a favorable
property for an efficient pair annihilation of neutrinos and
antineutrinos near the rotational axis.

\subsection{Energetics} 

\subsubsection{General feature}

Figure~\ref{fig2} plots the evolution of the total rest-mass energy
($M_*$), kinetic energy ($T_{\rm kin}$), internal energy ($E_{\rm
int}$), thermal energy ($E_{\rm th}$), and electromagnetic energy
($E_{\rm B}$), for which the integrations (\ref{eq5.1})--(\ref{eq5.5})
are performed for the region outside the apparent horizon, as well as
an averaged value of the specific internal energy defined by $E_{\rm
int}/M_*$ for models 3a0m2 (left) and 6a0m4 (right).  The left panel
of Fig.~\ref{fig2} quantitatively captures the features for the
evolution of the system described in Fig.~\ref{fig1}: In the first
$\sim 3$~ms, the electromagnetic energy increases due to the magnetic
winding and MRI. The electromagnetic energy increases to $\sim 10\%$
of the internal energy and $\sim 1\%$ of the kinetic
energy. Eventually, the increase of the electromagnetic energy
saturates, and then, the outflow of the material from the accretion
torus and the accretion of the material into the black hole are
activated.  Associated with these dynamical processes, the shock
heating is enhanced and the thermal energy steeply increases at $t
\sim 3$~ms. The averaged value of the specific internal energy
increases to $\langle \varepsilon \rangle
\sim 0.003$--0.004, implying that the averaged matter temperature is $\sim
2$--3~MeV. For $t \agt 10$~ms, the system relaxes to a quasisteady
accretion phase, during which the rest mass of the torus gradually decreases.
Associated with this decrease, $T_{\rm kin}$, $E_{\rm int}$, and
$E_{\rm B}$ also decrease gradually. On the other hand, $\langle
\varepsilon \rangle$ is approximately constant.  All these
quantitative features approximately hold for any model considered in
this paper. This fact is observed by the comparison between the left and
right panels of Fig.~\ref{fig2}.

\begin{figure}[t]
\begin{center}
\epsfxsize=2.5in
\leavevmode
(a)\epsffile{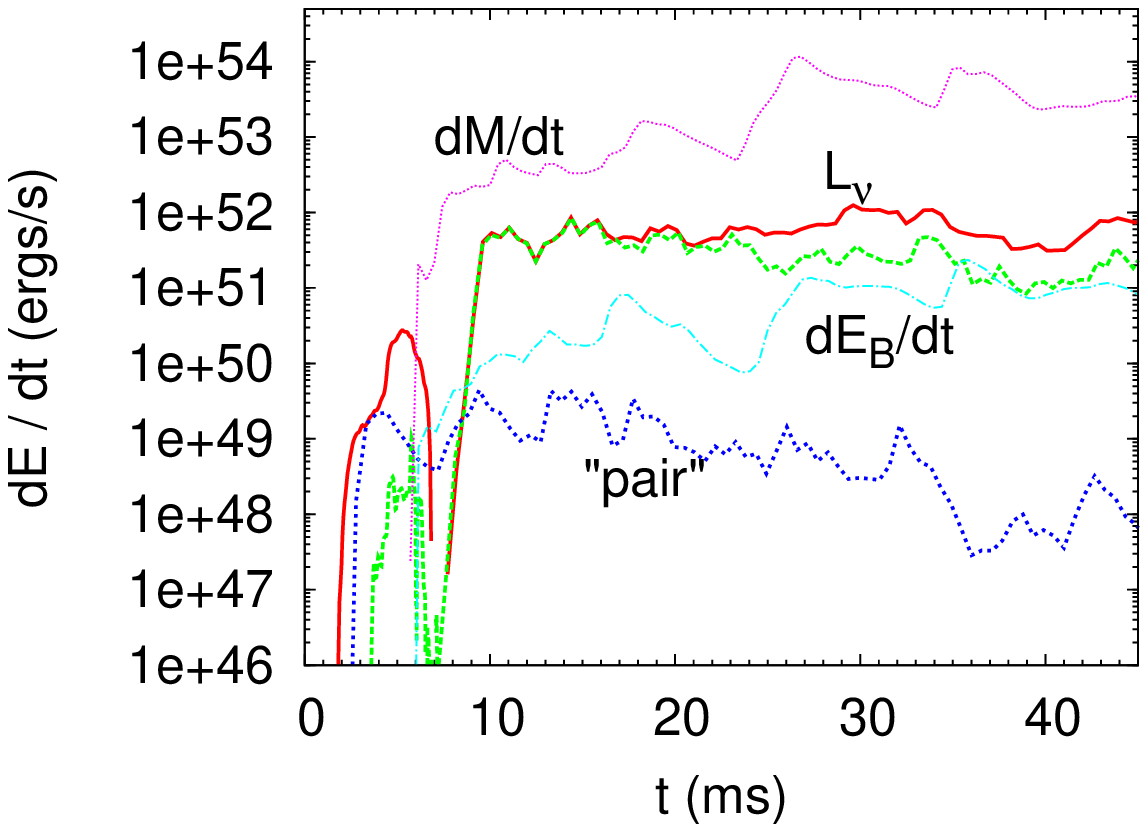}
\epsfxsize=2.5in
\leavevmode
~~(b)\epsffile{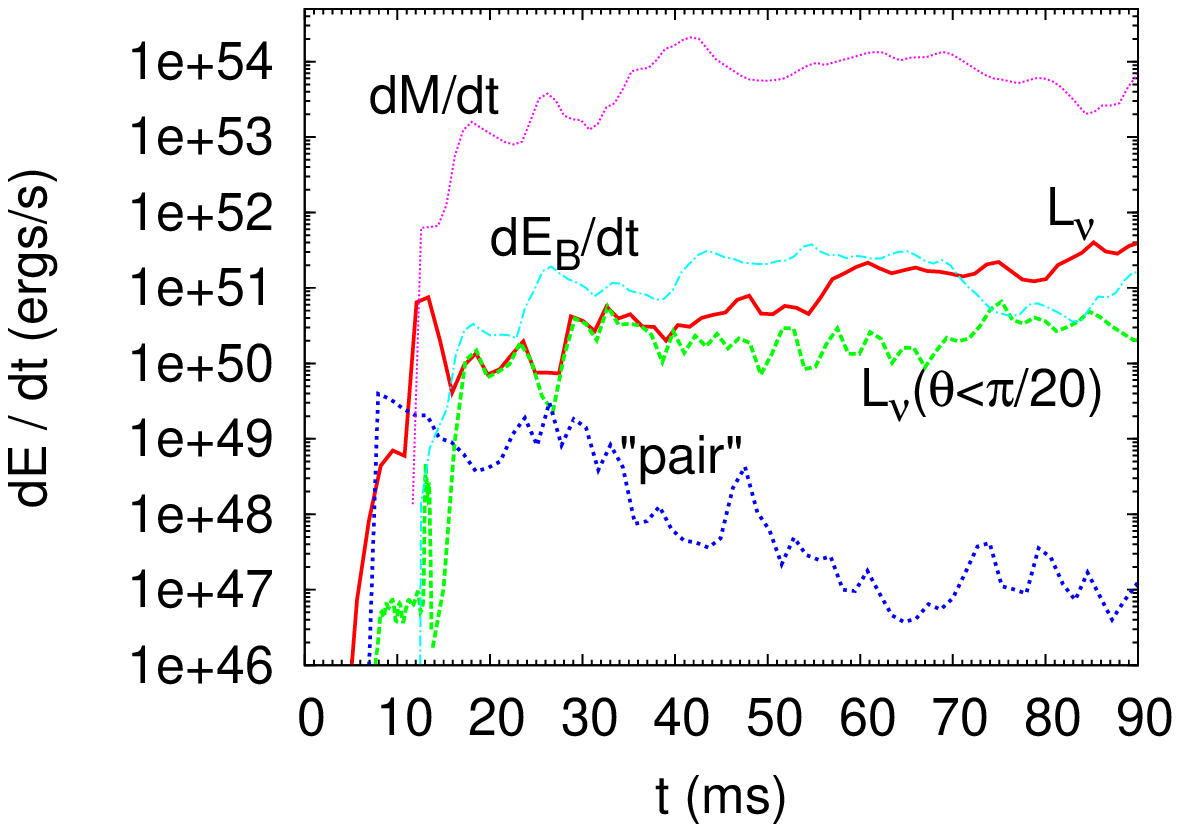}\\
\epsfxsize=2.5in
\leavevmode
(c)\epsffile{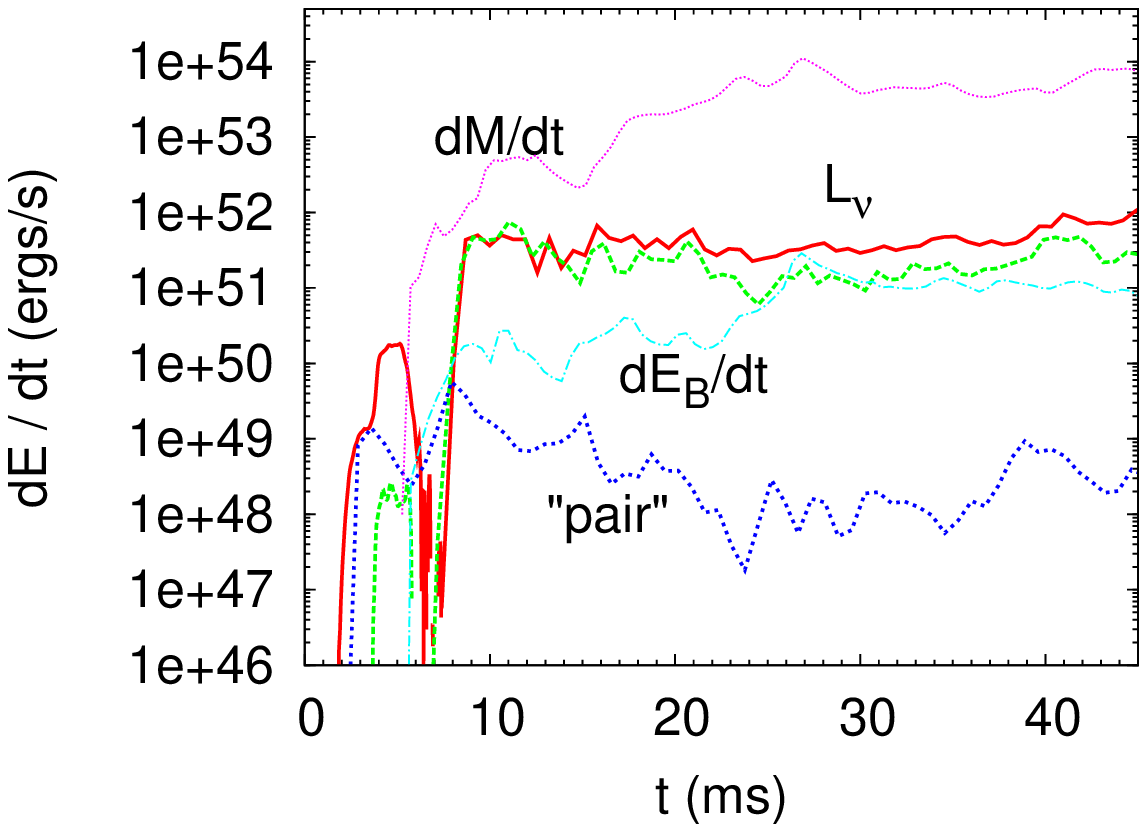}
\epsfxsize=2.5in
\leavevmode
~~(d)\epsffile{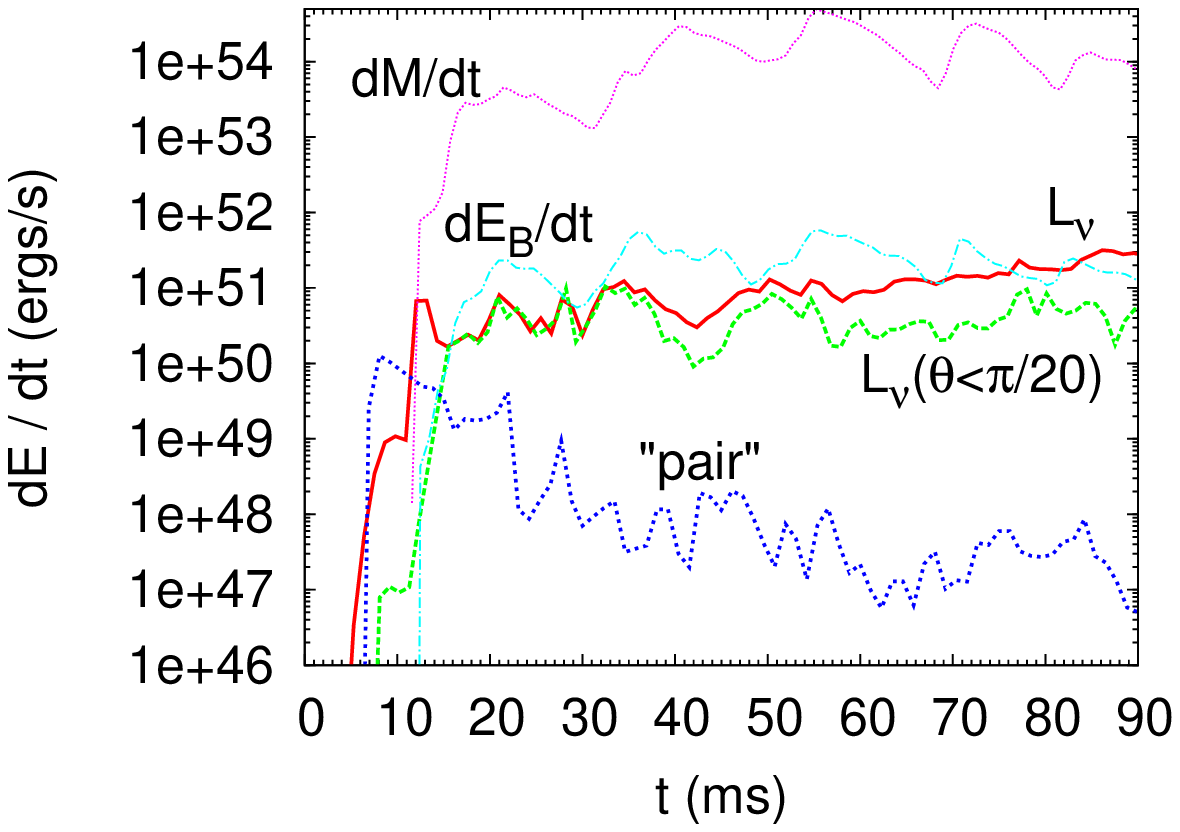}
\vspace{-3mm}
\caption{
Matter energy ejection rate ($dM/dt$), electromagnetic luminosity
($dE_{\rm B}/dt$), neutrino luminosity ($L_{\nu}$), and an approximate
estimate for the pair annihilation luminosity (``pair'') (a) for model
3a0m2, (b) for model 6a0m4, (c) for model 3a0m1, and (d) for model
6a0m6. The neutrino luminosity measured for the opening angle around
the rotational axis of $\theta \leq \pi/20$ is also plotted 
(the curves plotted in the vicinity of $L_{\nu}$).
\label{fig3}}
\end{center}
\end{figure}

%%% NO BZ luminosity

\subsubsection{Dependence on BH mass}

Figure~\ref{fig3} plots the evolution of the matter ejection rate
($dM/dt$), electromagnetic luminosity ($dE_{\rm B}/dt$), neutrino
luminosity ($L_{\nu}$), and an approximate estimate for the pair
annihilation luminosity by Eq.~(\ref{eq5.13}) for models (a) 3a0m2 (b)
6a0m4, (c) 3a0m1, and (d) 6a0m6. Because of the MHD
activity induced by the magnetic winding and MRI, the material in the
torus is ejected outward. The typical mass ejection rate is $\sim
0.1M_{\odot}$/s for $M_{\rm BH}=3M_{\odot}$ and $\sim 0.3M_{\odot}$/s
for $M_{\rm BH}=6M_{\odot}$ in our employed models. After the
activity of the torus turns on, the neutrino and electromagnetic
luminosities are also enhanced, and these luminosities eventually
relax to quasisteady values after the torus relaxes to a quasisteady
state. 

The typical neutrino luminosity is $\sim 10^{52}$~ergs/s for
$M_{\rm BH}=3M_{\odot}$ and $\sim 10^{51}$~ergs/s for 
$M_{\rm BH}=6M_{\odot}$ in our models. Thus, it is higher for the lower black
hole mass. By contrast, the dependence of the electromagnetic
luminosity, $\sim 10^{51}$~ergs/s, on the black hole mass is not as
strong as that of the neutrino luminosity. The electromagnetic
luminosity has a positive correlation with $dM/dt$ as $dE_{\rm
B}/dt=C_{\rm EM}dM/dt$.  The weak dependence of the ratio coefficient
$C_{\rm EM}\sim 3 \times 10^{-2}$ indicates that the mass ejection is
driven by the electromagnetic power.
%%%
An important point indicated in this work is that for
the lower-mass black-hole system, the neutrino emission is the primary
dissipation mechanism, while for the higher-mass case, the
electromagnetic wave emission plays the primary role.
We do not fully take into account the relevant 
microphysical processes and the results presented in this paper are
approximate ones. However, it is strongly likely that a critical mass, which
divides the dominant dissipation mechanism, exists.
The reason for this mass dependence will be explained as follows.

The weak dependence of $dE_{\rm B}/dt$ on the black hole mass may be
explained assuming that it is proportional to $B^2 R^3 \Omega$ as in
the Blandford-Payne mechanism~\cite{BP,meier}.  Here, $B$, $R$, and
$\Omega$ are the typical magnetic-field strength, radius, and angular
velocity of the torus. We note that for the models employed in this
paper, the black hole spin is zero (or small, as a result of
accretion), and hence, the Blandford-Znajek mechanism is not
relevant~\cite{BZ}. The present models also have the parameters
$R/M_{\rm BH} \sim 10$ and $\Omega \propto M_{\rm BH}^{-1}$
universally.  For the saturated magnetic field strength, $B^2$ is
approximately proportional to the gas pressure $P \propto
\rho c_s^2 \sim \rho T$ where $c_s$ is the sound velocity. Here, $T$
depends only weakly on the black hole mass, and we have an approximate
relation $\rho \propto M_{\rm BH}^{-2}$ for a given mass and radius of
the torus.  Thus, $B^2$ should be proportional to $M_{\rm BH}^{-2}T$,
and $B^2 R^3 \Omega
\propto M_{\rm BH}^{0}T$: $dE_{\rm B}/dt$ should depend only weakly on
the black hole mass.  This agrees with the present numerical results.

The dependence of the neutrino luminosity on the black hole mass may
be estimated in the following manner. Suppose that the mass accretion
rate of a torus into the black hole is $\dot M_{\rm in}$. Then, the
accretion luminosity $L_{\rm acc}$ is approximately written as $M_{\rm
BH} \dot M_{\rm in}/R$. Because $M_{\rm BH}/R$ is $\sim 0.1$ depending
weakly on the black hole mass, $L_{\rm acc} \propto \dot M_{\rm in}$.
For a given disk mass, $\dot M_{\rm in} \sim M_{\rm disk}/\tau$ where
$\tau$ is the accretion time scale which should be proportional to
$M_{\rm BH}$. Thus, $L_{\rm acc} \propto M_{\rm BH}^{-1}$.  If the
neutrino luminosity is proportional to $L_{\rm acc}$, it should be
proportional to $M_{\rm BH}^{-1}$: For the smaller black hole mass,
the neutrino luminosity should be larger.  This agrees qualitatively
with the present numerical results.

\subsubsection{Neutrino pair annihilation luminosity}

In Fig.~\ref{fig3}, the neutrino luminosity measured for the angle
$\theta \leq \pi/20$ is also plotted. Comparing this with the total
neutrino luminosity shows that neutrinos are emitted substantially
toward the outward direction along the rotation axis. This is likely
to come from the facts that the inner region of the torus has the
maximum temperature and that a funnel structure is formed after the
magnetic instabilities turn on.  This collimation effect enhances the
neutrino radiation density near the axis, and as a result, the 
efficiency of the pair annihilation of neutrinos and antineutrinos
will be also enhanced.

The pair annihilation luminosity depends most strongly on the black
hole mass. The reason is that it is proportional to the square of the
neutrino luminosity. For the binary-neutron-star model with $M_{\rm
BH}=3M_{\odot}$, the typical magnitude is $10^{48}$--$10^{49}$~ergs/s
whereas for the black-hole-neutron-star-binary model with $M_{\rm
BH}=6M_{\odot}$, it is $\sim 10^{47}$~ergs/s. For a hypothetical
duration of the neutrino emission $\sim 100$~ms for $M_{\rm
BH}=3M_{\odot}$, the total energy deposited could be $\sim
10^{48}$~ergs. This value could explain the total energy of relatively
weak SGRB.  By contrast, for $M_{\rm BH}=6M_{\odot}$, the total energy
could be at most $10^{47}$~ergs even for the time duration of 1~s.
This is too low to explain the observed SGRB. However, a word of the
note is appropriate here. First, we do not take into account some of
important neutrino emission processes and realistic microphysics in
this work. The neutrino luminosity could be higher than that estimated
in this work in reality. If so, the pair-annihilation luminosity could
be higher than that presented in this paper. Second, we do not
consider the spin of black holes, although a black hole of a
relatively high spin is likely to be formed for the merger of binary
neutron stars and black hole-neutron star binaries.  For such a
high-spin black hole, the neutrino luminosity could be by one order of
magnitude larger than that presented in this
paper~\cite{chen,SST07}. This suggests that the pair-annihilation
luminosity could be by two order of magnitude larger for the case that
a black hole has a high spin. More detailed study taking into account
the detailed microphysics and high black hole spin is the issue for
our subsequent work. 

\section{Summary}

We present our first numerical result of the GRRMHD simulation for
black hole-torus systems in the framework of full general
relativity. Taking into account the latest results of the
numerical-relativity simulation for black hole-neutron star binaries
and binary neutron stars, we evolve the systems of $M_{\rm
BH}=3M_{\odot}$ or $6M_{\odot}$ and of the torus mass
$0.14$--$0.38M_{\odot}$ with the maximum density of the torus $5
\times 10^{11}$--$3\times 10^{12}~{\rm g/cm^3}$. We incorporate 
some of neutrino processes (absorption and emission via the
interaction with free nucleons) in an approximate manner. The
following is the summary for the results obtained in this paper:
\begin{itemize}
\item The typical order of the neutrino luminosity is $10^{52}$~ergs/s 
for $M_{\rm BH}=3M_{\odot}$ and $10^{51}$~ergs/s for $M_{\rm
BH}=6M_{\odot}$. Because we do not take into account some 
of important processes of the neutrino emission, these values 
may be considered as a lower bound for the neutrino luminosity. 
%%%%%%%%%%%%%%%%%%
\item Neutrinos are dominantly emitted toward the outward direction 
along the rotation axis of the accretion torus. The reason is that
after MHD instabilities set in, a high-temperature region appears in
the inner edge of the accretion torus and also a funnel structure,
which enhances the emission toward the direction along the rotational
axis, is formed.
%%%%%%%%%%%%%%%%%%
\item The order of the electromagnetic luminosity is $10^{51}$~ergs/s 
and it depends only weakly on the black hole mass in our present
models. Thus, for $M_{\rm BH}=3M_{\odot}$, the neutrino luminosity is
larger than the electromagnetic luminosity, while for $M_{\rm
BH}=6M_{\odot}$, two luminosities are comparable. The
order-of-magnitude estimate supports this result, and hence, it is
likely that there is a critical black hole mass above which the
neutrino luminosity is smaller than the electromagnetic luminosity.
%%%%%%%%%%%%%%%%%%
\item A very approximate estimate for the total pair annihilation 
rate of neutrino-antineutrino pairs suggests that the order of the
magnitude of the luminosity associated with this process is
$10^{49}$~ergs/s for $M_{\odot}=3M_{\odot}$ and $10^{47}$~ergs/s for
$M_{\odot}=6M_{\odot}$. Again, these values may be considered as a
lower bound because we do not incorporate some of important processes
of the neutrino emission nor the black hole spin; the value may be
underestimated by a factor of 10--100. Taking into account this fact,
the present results suggest that SGRB with relatively low energy may
be driven from the remnants formed from the merger of black
hole-neutron star binaries and binary neutron stars, respectively.
\end{itemize}

\vskip 3mm
\begin{center}
{\bf Acknowledgments}
\end{center}
\vskip 3mm

This work was supported by Grant-in-Aid for Scientific Research
(21018008, 21105511, 21340051, 23740160), by Grant-in-Aid for
Scientific Research on Innovative Area (20105004), and by HPCI
Strategic Program of Japanese MEXT. 

\appendix

\section{Radiation field in the optically thick spherical flow}

As shown in Refs.~\citen{AS,SKSS11}, it is possible to derive an
approximate solution of the radiation field in the optically thick
limit using the diffusion approximation. We here describe the analytic
solution of the Bondi flow in the framework of radiation
hydrodynamics, i.e., an accretion flow of the matter and radiation in
the Schwarzschild background~\cite{ST}.  We also show that our
numerical code can generate the analytic solutions for the radiation
flows $({\cal J}, {\cal H}_{\alpha})$ and $({\cal E}, {\cal F}_i)$.

We choose the line element in the Kerr-Schild coordinates,
\beqn
ds^2=-\Big(1 - {2M \over r}\Big)d{\bar t}^2 
+{4M \over r} d{\bar t}dr + \Big(1 + {2M \over
r}\Big)dr^2+r^2 (d\theta^2 + \sin^2\theta d\varphi^2), 
\label{KS}
\eeqn
where $M$ and $r$ are the gravitational mass and areal coordinate.  In
this metric, the coordinate singularity at $r=2M$ does not give any
messy problem. 

First, we consider the optically thick limit. In this case, 
the total stress-energy tensor is written as
\beqn
T^{\rm tot}_{\mu\nu}=\Big(\rho h + {4 \over 3}{\cal J}^{\rm eq}\Big)u_{\mu}u_{\nu}
+\Big(P + {1 \over 3}{\cal J}^{\rm eq}\Big) g_{\mu\nu}. 
\eeqn
Here we do not consider the magnetic field for simplicity.  However,
it is straightforward to take into account a purely radial magnetic
field~\cite{MG} and our code was already shown to reproduce such 
a solution~\cite{SS05}. 

In the following, we assume that the radiation is composed of photons,
writing ${\cal J}^{\rm eq}=a_{\rm r} T^4$, and consider the case that the EOS
of the matter is described by $P=K \rho^{4/3}=\rho \varep/3$, i.e.,
the $\Gamma=4/3$ ideal fluid. Then for the ideal gas, $\varep=3k_{\rm
B}T/m$ where $m$ is the mass of the gas particle.  In this EOS,
${\cal J}^{\rm eq}$ is written as $3K'\rho^{4/3}$ where $K'=a_{\rm
r}(mK/k_{\rm B})^4/3$. Then, the total stress-energy tensor is written
by
\beqn
T^{\rm tot}_{\mu\nu}=\Big(\rho + 4(K+K')\rho^{4/3}\Big)u_{\mu}u_{\nu}
+(K+K')\rho^{4/3} g_{\mu\nu}. 
\eeqn
Namely, the stress-energy tensor is the same as that for the effective
EOS, $P_{\rm eff}=(K+K')\rho^{4/3}$.  This implies that it is easy to
generate a Bondi solution, i.e., $\rho$, $u^r$, and $P_{\rm eff}$ as
functions of $r$ for a given value of $K+K'$.  If the ratio of $K'/K$
is determined, we also obtain $P$ and ${\cal J}^{\rm eq}$ as functions of
$r$.

In the optically thick medium, the correction of the
radiation fields in our notation is written as~\cite{SKSS11}
\beqn
{\cal J}={\cal J}^{\rm eq}+l {\cal J}^{(1)}~~{\rm and}~~{\cal H}_i=l H^{(1)}_i,
\eeqn
where $l$ is the mean free path of the radiation, $\kappa^{-1}$, which
is smaller than the typical size of the system (i.e., $l \ll M$), and
\beqn
&& {\cal J}^{(1)}=-u^{\mu} \nabla_{\mu} {\cal J}^{\rm eq}
-{4 \over 3}{\cal J}^{\rm eq}\nabla_{\mu} u^{\mu},\\
&& {\cal H}^{(1)}_{\alpha}=-{1 \over 3}h_{\alpha}^{~\beta} \nabla_{\beta} 
{\cal J}^{\rm eq}-{4 \over 3}{\cal J}^{\rm eq} u^{\beta} \nabla_{\beta} u_{\alpha}.
\eeqn
Here, ${\cal J}^{\rm eq}$ is written as $3K'P/K$. 
Using this relation, we find
\beqn
{\cal J}^{(1)}=0,~~~{\cal H}^{(1)}_r=-{1+u_r u^r \over 3} {\cal J}^{\rm eq}_{,r}
-{4 \over 3} {\cal J}^{\rm eq}\Big(u^r u_{r,r}+{M \over r^2}(u^t+u^r)^2\Big), 
\eeqn
and hence, 
\beqn
&&{\cal J}={\cal J}^{\rm eq} + O(l^2),\label{eqa8}\\
&&{\cal H}_i=l {\cal H}^{(1)}_r \gamma^r_{~i} + O(l^2).\label{eqa9}
\eeqn
In the numerical simulation, we evolve the variables ${\cal E}_{0}$ and
$({\cal F}_{0})_{i}$. Then, we derive ${\cal J}$ and ${\cal H}_i$ using the relation for the
optically thick case as
\beqn
&& {\cal J}={3 \over 2w^2 + 1}\Big[(2w^2-1) {\cal E}_{0} - 2w ({\cal F}_{0})_i V^i \Big],\\
&& {\cal H}_i=w^{-1} ({\cal F}_{0})_i + 
{1 \over w(2w^2 + 1)}\Big[-4w^3 {\cal E}_{0} + (4w^2+1) u_i ({\cal F}_{0})_j V^j \Big]. 
\eeqn
%%\beqn
%%&&e={4w^2 - 1 \over 3} J^{\rm eq}+2 l H^{(1)}_i u_j \gamma^{ij},\\
%%&&f_i={4w \over 3}u_i J^{\rm eq}+l\Big(
%%w H^{(1)}_i +{u_i \over w}H^{(1)}_j u_k \gamma^{jk}\Big). 
%%\eeqn

The test simulations were performed for a Bondi flow solution with a
critical radius $12M$ and $dM/dt=1$. We set $\kappa$ to be constant
and choose $\kappa=1$, 10, 30, and 100. The computational domain with
$[0:50M]$ for both $x$ and $z$ is covered by the uniform grid with the
grid number $N=250$, 500, and 1000. The simulations were always
performed for the time duration of $50M$ because for such a duration,
a relaxed converged solution is obtained. The results shown in the
following are the snapshots at $t=50M$. The time step of the
simulation was fixed to be $0.5\Delta x$ where $\Delta x=50M/N$.

\begin{figure}[t]
\begin{center}
\epsfxsize=2.6in
\leavevmode
\epsffile{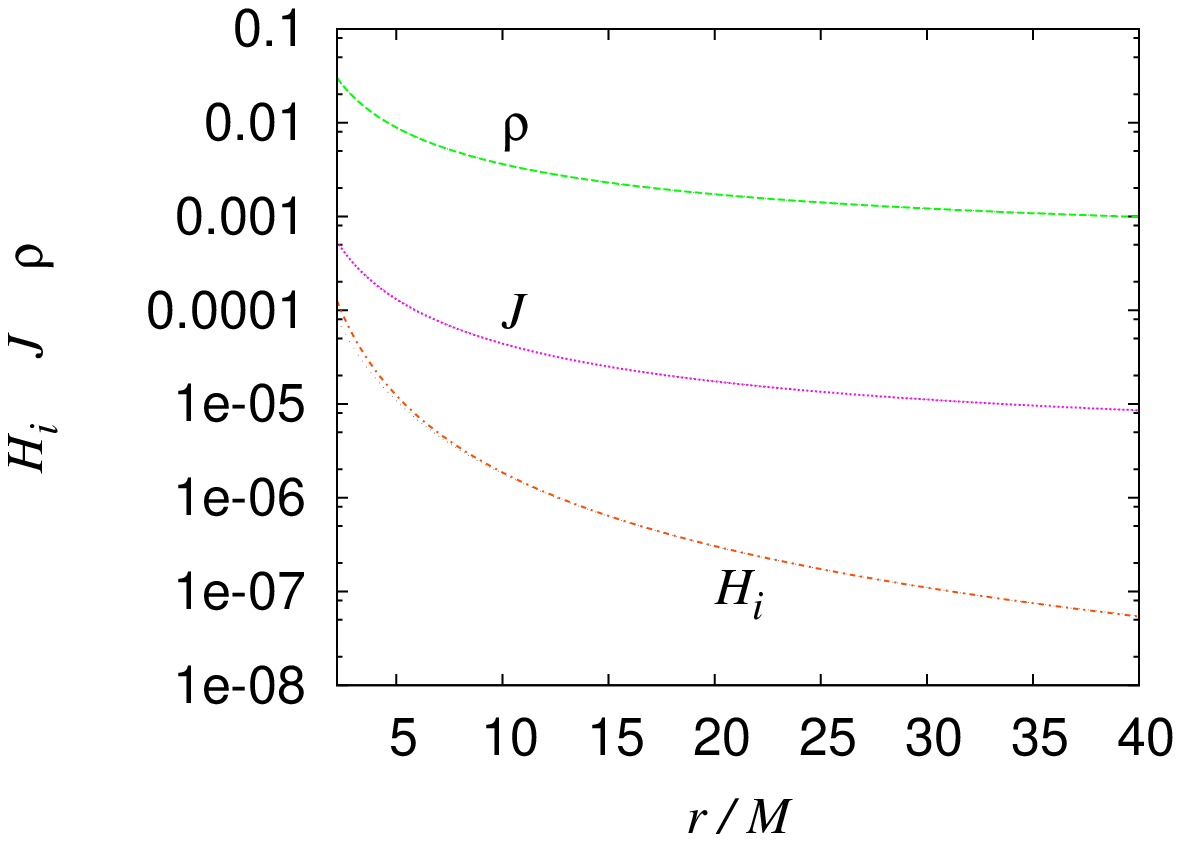}
\epsfxsize=2.6in
\leavevmode
~~~\epsffile{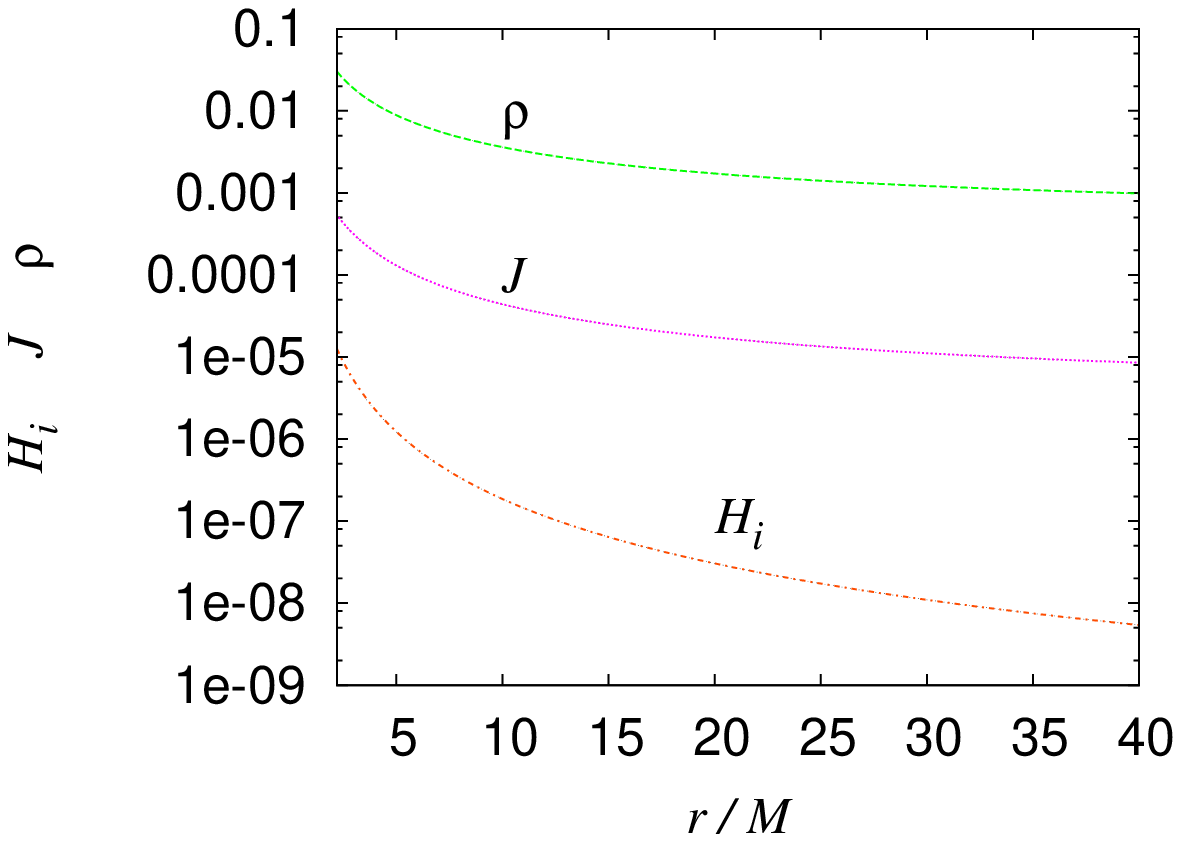}
\vspace{-1mm}
\caption{$\rho$, ${\cal J}$, and ${\cal H}_i$ as functions of $r/M$ 
for $\kappa=1$ (left) and 10 (right) and for $N=1000$.  The points and
dashed curves denote the numerical results and analytic solutions.
\label{figa1}}
\end{center}
\end{figure}

\begin{figure}[t]
\begin{center}
\epsfxsize=2.6in
\leavevmode
\epsffile{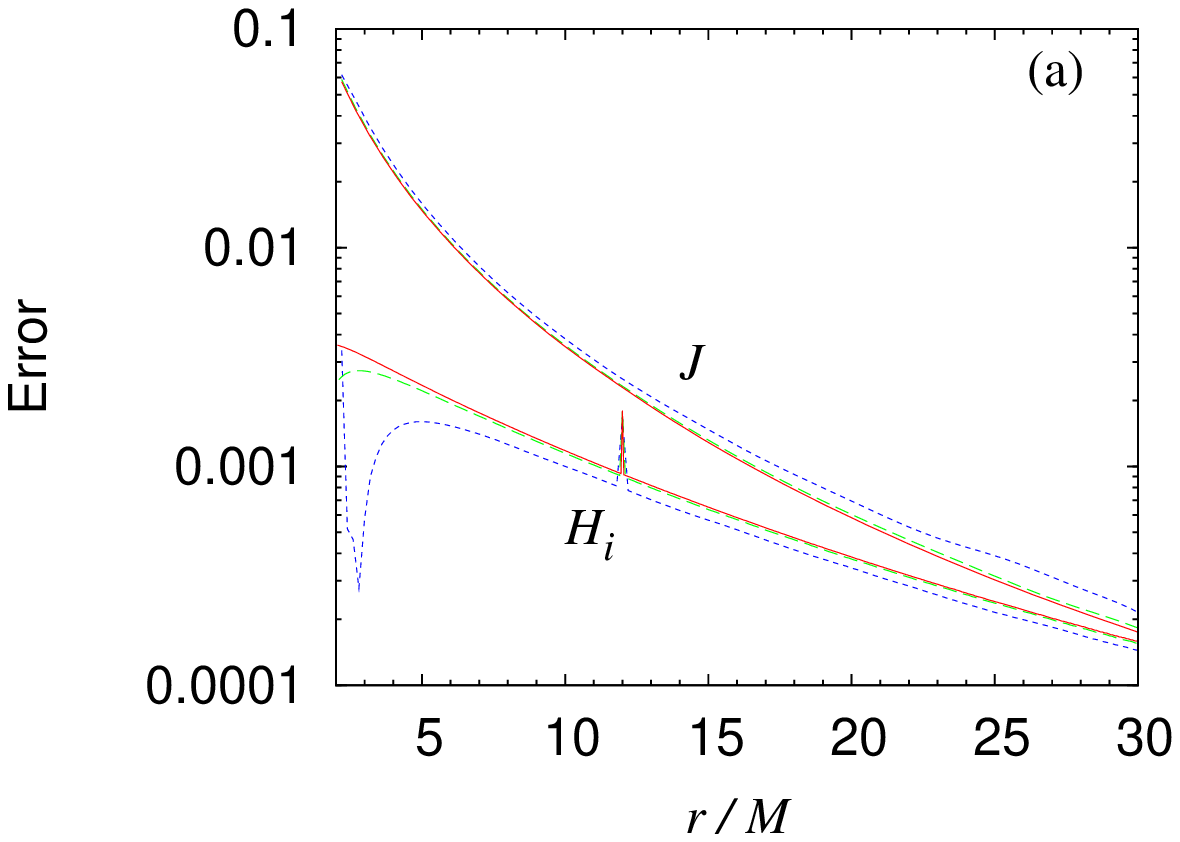}
\epsfxsize=2.6in
\leavevmode
\epsffile{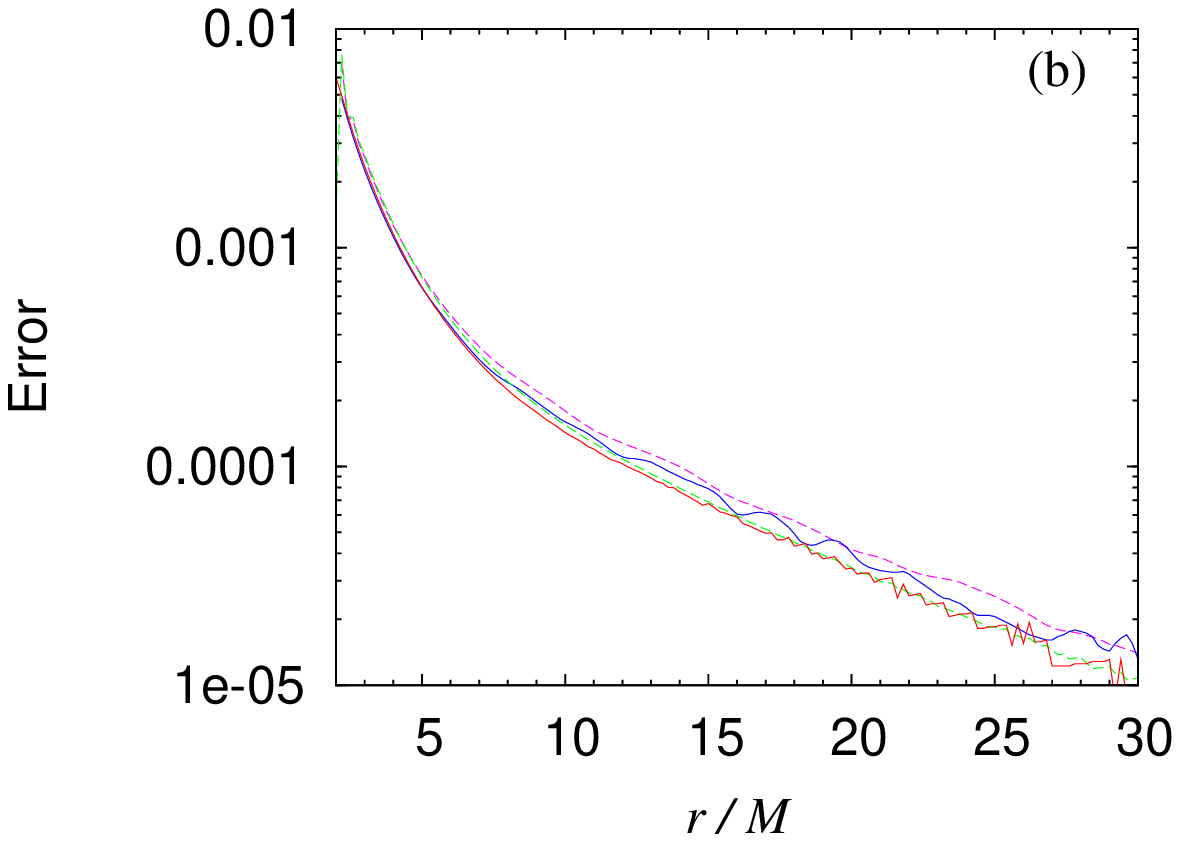}
\vspace{-1mm}
\caption{(a)The relative error of ${\cal J}$ and ${\cal H}_i$ as functions of $r/M$ 
for $\kappa=10$. The solid, dashed, and dotted curves are the results
for $N=1000$, 500, and 250, respectively.  (b) Comparison of
$5[{\cal J}(N=500)-{\cal J}(N=1000)]$ (dashed curves) and $[{\cal J}(N=250)-{\cal J}(N=1000)]$
(solid curves) for $\kappa=1$ (upper) and 10 (lower curves).
\label{figa2}}
\end{center}
\end{figure}

Figure \ref{figa1} plots $\rho$, ${\cal J}$, and ${\cal H}_i$ as functions of $r/M$
for $\kappa=1$ and 10, and $N=1000$.  The points and solid curves are
the numerical results and approximate analytic solutions valid up to
$O(l)$. The data are extracted along the $x$-axis.  This figure shows
that the numerical computation reproduces the analytic solutions quite
accurately for $\kappa=10$.  For $\kappa=1$, the diffusion
approximation does not accurately hold near the black hole horizon,
and $H_i$ for the numerical solution does not agree with the analytic
solution. However, for the outer region, the diffusion approximation
still works.  It should be noted that the magnitude of ${\cal H}_i$ is
proportional to $\kappa^{-1}$ for $\kappa > 1$. This is also found
from this figure.

Next, we focus on the convergence of the numerical solutions.  Figure
\ref{figa2}(a) plots the relative errors for ${\cal J}$ and ${\cal H}_i$ along the
$x$-axis. Here, the relative error of a quantity ${\cal Q}$ is defined by
\beqn
{\rm Error}=
\Big|{{\cal Q}_{\rm numerical~solution} \over {\cal Q}_{\rm analytic~solution}}-1 \Big|,
\eeqn
and the analytic solution is that denoted by Eqs.~(\ref{eqa8}) and
(\ref{eqa9}). Figure \ref{figa2}(a) shows that the error does not
decrease with improving the grid resolution. The reason is that the
accuracy of the analytic solutions is not good enough due to the
neglection of the term of order $O(l^2)$.

To confirm that the numerical solution shows the convergent 
behavior, we then compare 
\beqn
5 [{\cal Q}(N=500)-{\cal Q}(N=1000)]~~{\rm and}~~{\cal Q}(N=250)-{\cal Q}(N=1000),
\eeqn
where ${\cal Q}(N)$ denotes a numerical result with the grid number $N$.  If
the computation is second-order convergent, i.e., ${\cal Q}$ behaves as
${\cal Q}={\cal Q}(N \rightarrow \infty)+\Delta x^2 {\cal Q}_2$, these two quantities
should agree.  Figure \ref{figa2}(b) compares these two quantities for
${\cal J}$ and ${\cal H}_i$.  This indeed shows the agreement and that the numerical
results are second-order convergent. Note that the reason of the
violation of the second-order convergence for the error smaller than
$\sim 10^{-4}$ is that other error sources come into the play for such
a small error size. 

\begin{figure}[t]
\begin{center}
\epsfxsize=2.6in
\leavevmode
\epsffile{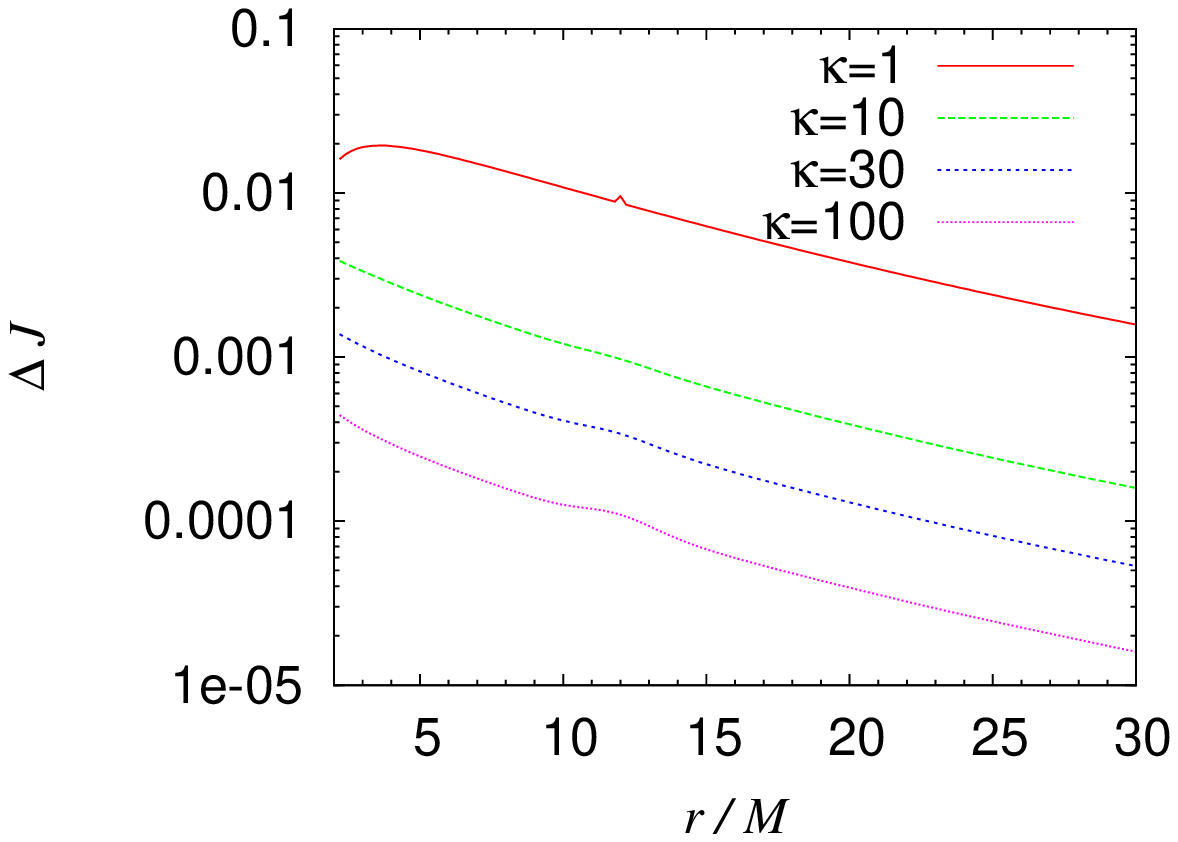}
\epsfxsize=2.6in
\leavevmode
~~\epsffile{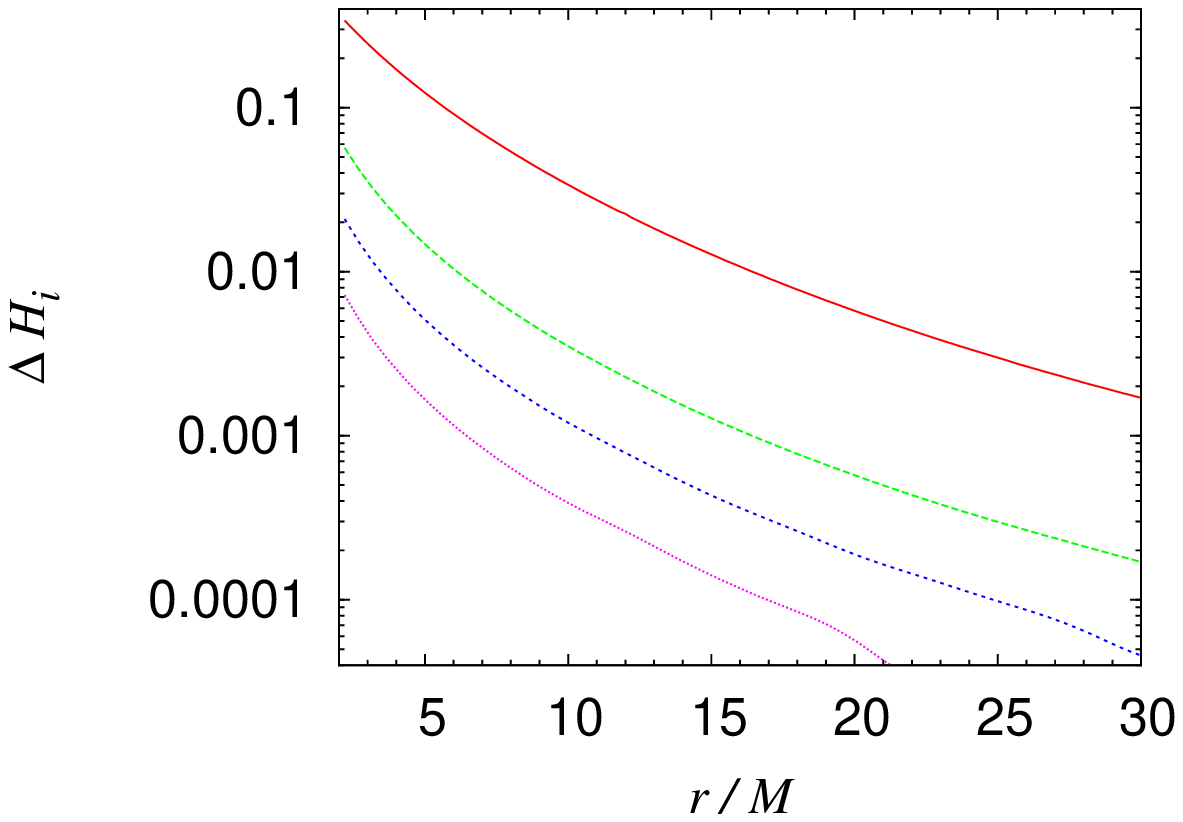}
\vspace{-1mm}
\caption{The relative error between the extrapolated 
results and analytic results of ${\cal J}$ and ${\cal H}_i$ as functions of $r/M$. 
The magnitude of the error should be proportional to $\kappa^{-1}$
and the figure shows that this is indeed the case. 
\label{figa3}}
\end{center}
\end{figure}

Because the numerical results obey the second-order convergent
behavior, it is possible to derive an exact solution for $N
\rightarrow \infty$ approximately. Such an ``exact'' solution is 
different from the approximate analytic solution in which 
the term of $O(l)$ is not taken into account. Thus the 
relative difference between the ``exact'' solution and 
the analytic solution should be proportional to $\kappa^{-1}$. 
To confirm that the numerical solution reproduces this fact, 
we plot the following relative error of ${\cal H}_r$ in Fig. \ref{figa3} 
for $\kappa=1$, 10, 30, and 100:  
\beqn
\Big|{{\cal Q}_{\rm ``exact''~solution} \over {\cal Q}_{\rm
analytic~solution}}-1\Big|. 
\eeqn
It is found that this error is approximately proportional to
$\kappa^{-1}$. This fact further validates our numerical solutions. 

\section{One dimensional special relativistic tests}

The University of Illinois at Urbana-Champaign (UIUC) group proposed
four one-dimensional test problems of the radiation hydrodynamics in
special relativity~\cite{UIUC}, for the optically thick limit (the
Eddington closure, ${\cal L}_{\mu\nu}={\cal J} h_{\mu\nu}/3$, is assumed). They
derived four semi-analytic stationary solutions for Riemann problems.
Then, they performed the following test: The derived semi-analytic
solutions are boosted by a Lorentz transformation and they confirmed
that their code can reproduce the boosted non-stationary solutions.
However, by this test, one can check only the fact that the code can
reproduce a coordinate translation of given stationary solutions. 

Subsequently, Zanotti et al. \cite{Zanotti}) proposed a more
nontrivial shock tube test in which the stationary solutions by the
UIUC group should be produced in a dynamical simulation: They
initially prepared a discontinuous initial condition and performed
simulations until the numerical data relaxes to a solution derived in
Ref.~\citen{UIUC}. Hence, we performed the test simulations in the
same manner as in Ref.~\citen{Zanotti}.  We confirmed that our code
can accurately produce these solutions as illustrated in the
following. 

In the test of Ref.~\citen{Zanotti}, the stationary solution is
numerically derived preparing the initial conditions shown in
Table~\ref{table5}. In these conditions, the discontinuity at $x=0$ is
present, but during the dynamical evolution, the matter and radiation
fields relax to a stationary solution after a sufficiently time
duration.  We performed the simulations until the stationarity was
reached; the time duration of the numerical simulations is 5000 for
the tests 1 and 2, 100 for the test 3, and 300 for the test 4.  The
numerical simulations were performed with the $\Gamma$-law EOS,
$P=(\Gamma-1)\rho \varepsilon$.  The opacity is assumed to have the
form $\kappa=\kappa_0 \rho$ where $\kappa_0$ is a given constant. The
mean free path defined by $1/\kappa$ is rather long $\agt 0.1$ in
these tests, although the Eddington closure is assumed.  The
temperature is simply determined by $P/\rho$ during the simulation.

In the tests 1 and 2, the radiation pressure is much smaller than the
gas pressure. Thus, these are the tests in which one checks whether
the radiation fields are accurately evolved in an approximately fixed 
dynamical fluid flow with shocks. In the tests 3 and 4, the radiation
pressure is as strong as or stronger than the gas pressure. These are
regarded as the tests by which one checks whether the code can
accurately derive the system in which both the gas and radiation play
an important role. 

The simulations were performed with the grid spacing $\Delta x=0.03$
for the computational domain $[-15:15]$ for the tests 1--3. For the
test 4, we set $\Delta x=0.02$ with the computational domain
$[-20:20]$. At the outer boundaries, we imposed the boundary condition
that the fluid and radiation states remain to be the initial states
which are identical with the semi-analytic solutions.  The time step
is set to be $0.5\Delta x$ for all the tests.

\begin{figure}[t]
\begin{center}
\epsfxsize=2.6in
\leavevmode
\epsffile{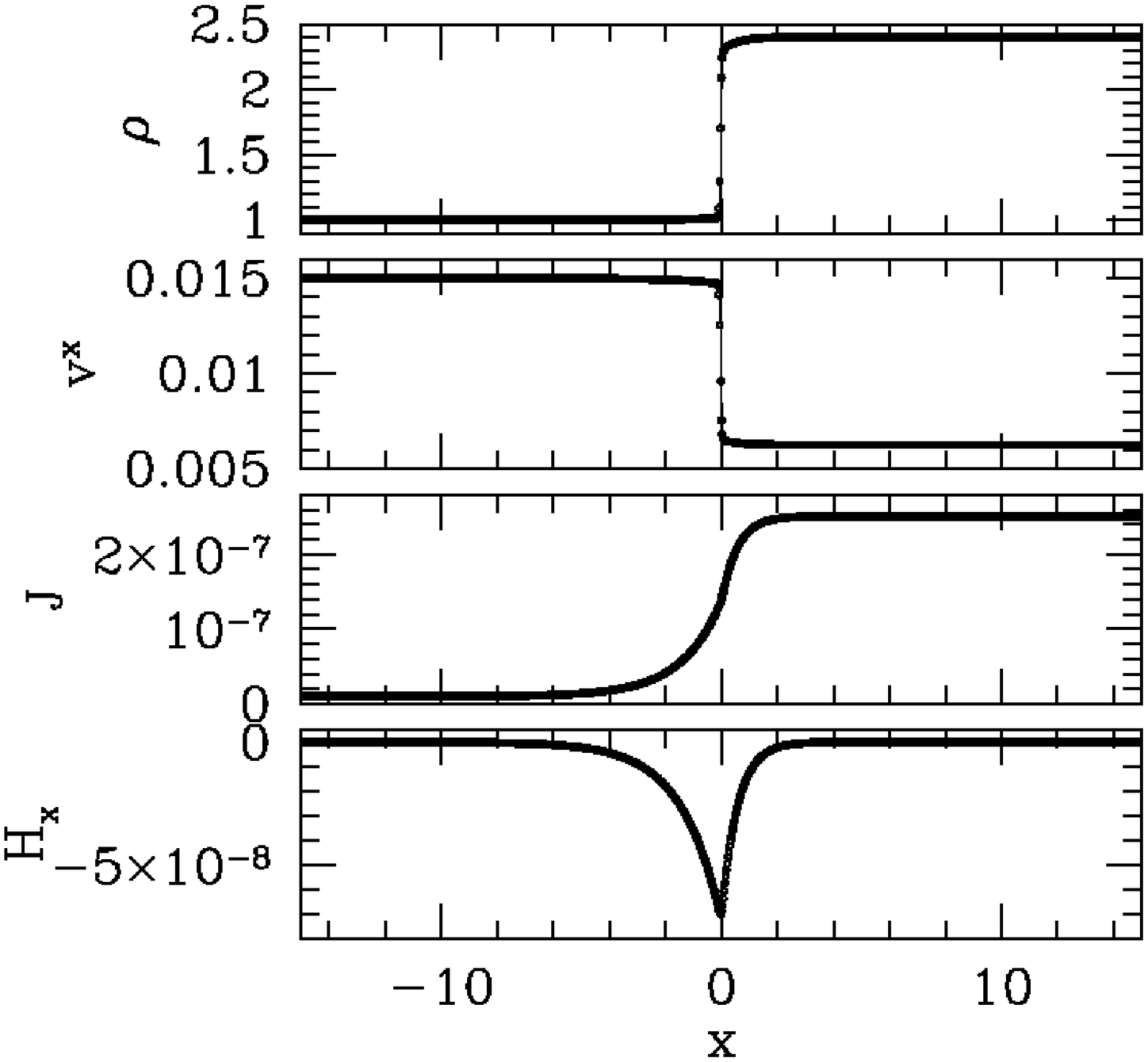}
\epsfxsize=2.6in
\leavevmode
~~\epsffile{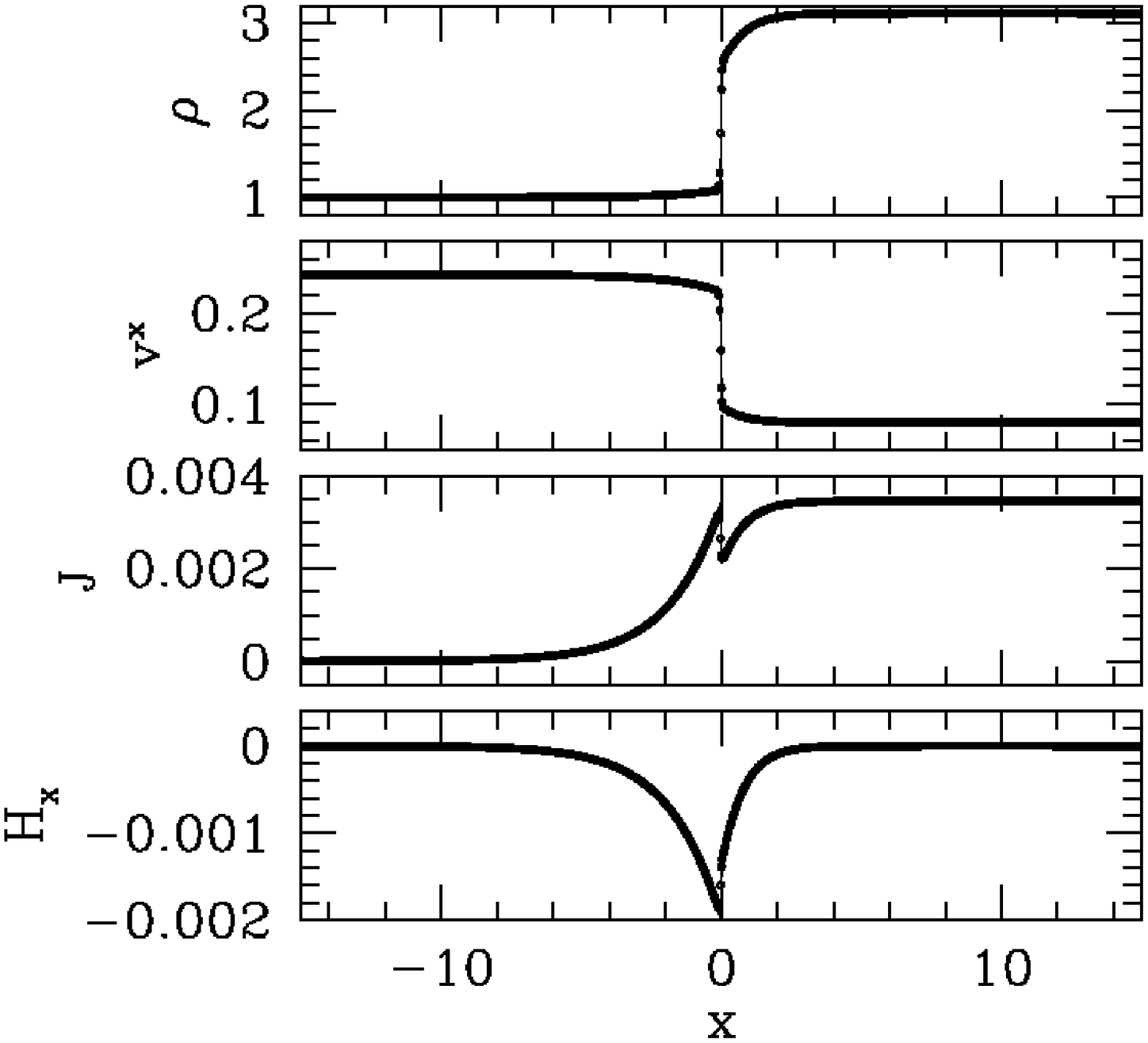} \\
\epsfxsize=2.6in
\leavevmode
\epsffile{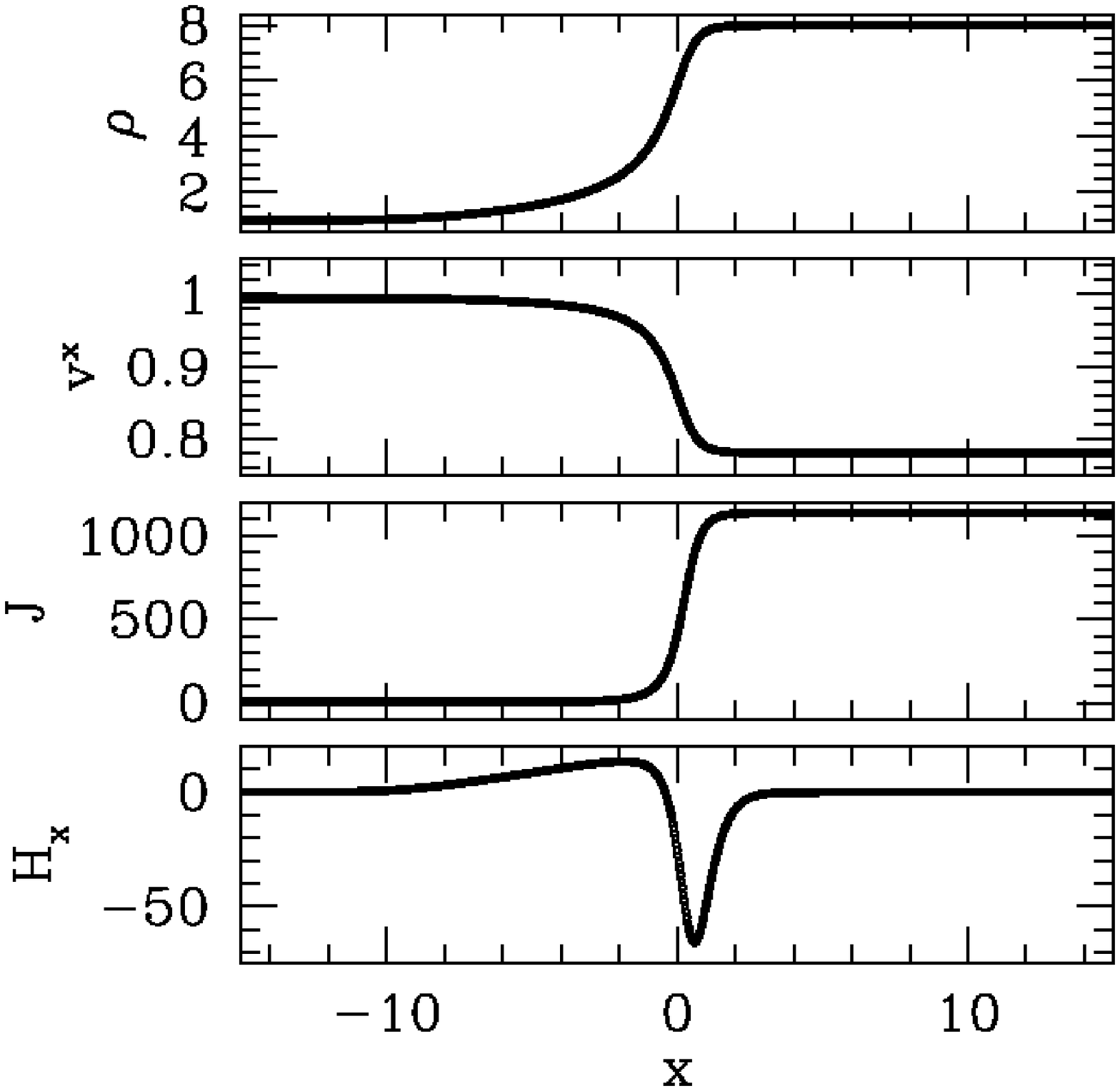} 
\epsfxsize=2.6in
\leavevmode
~~\epsffile{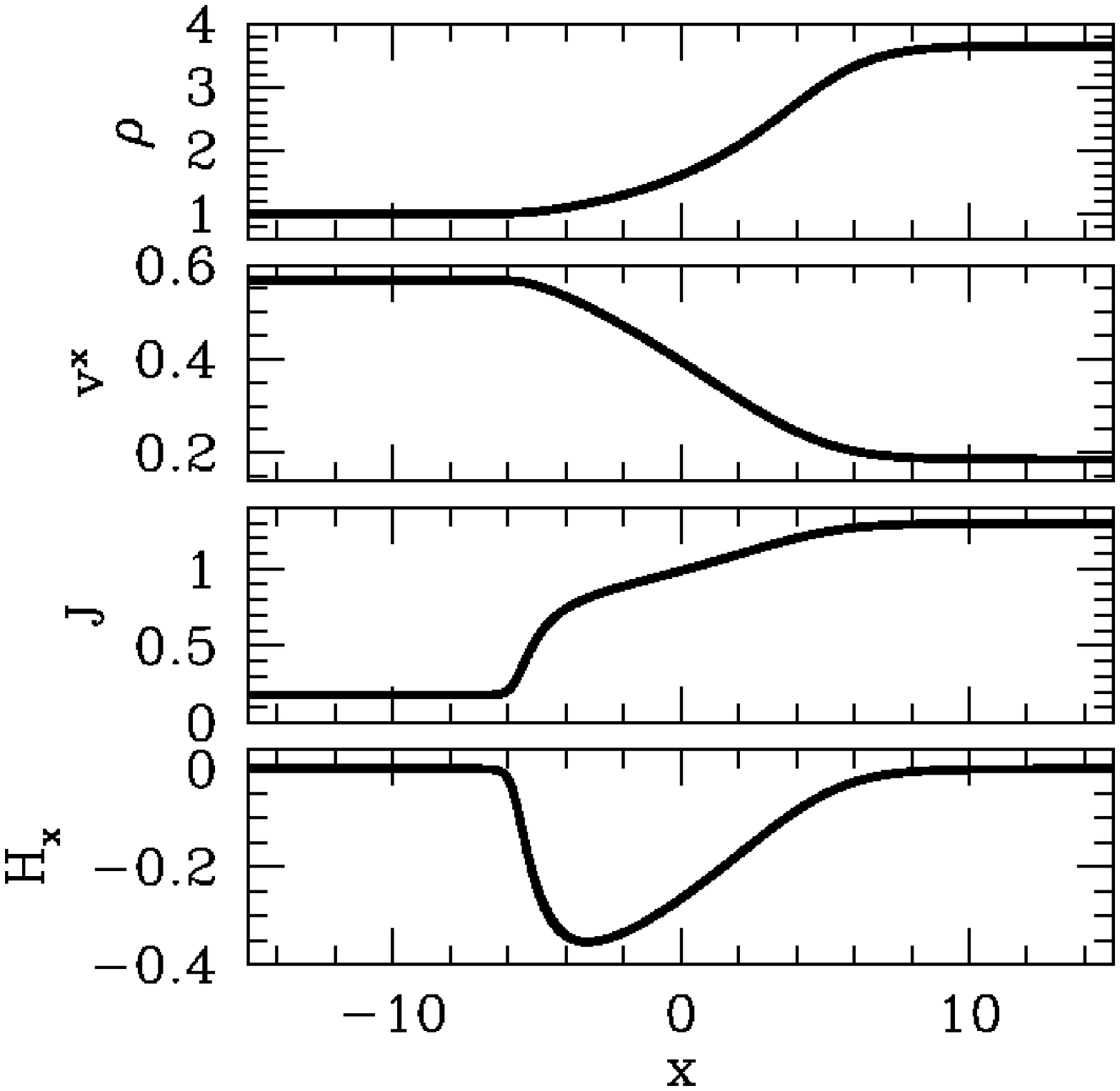} 
\vspace{-1mm}
\caption{Numerical results for one-dimensional 
test problems in special relativity (test 1: top left, test 2: top
right, test 3: bottom left, and test 4: bottom right). The solid
curves and open circles denote the semi-analytic solutions and
numerical results, respectively. 
\label{figa4}}
\end{center}
\end{figure}

Figure~\ref{figa4} plots the numerical results for all the four test
simulations. The solid curves and open circles denote the
semi-analytic solutions and numerical results, which overlap with each
other and cannot be distinguished in these figures. Thus, 
Fig.~\ref{figa4} shows that for all the four test simulations, the
numerical computation satisfactorily reproduces the semi-analytic
solutions, as in Refs.~\citen{Zanotti}. This shows an evidence that
our numerical code is reliable. 

Before closing this section, we note the following fact for the test
3.  In this test, the stationary solution for the region $-10 \alt x
\alt 0$ satisfies ${\cal J}^2 < {\cal H}_{\mu}{\cal H}^{\mu}$ and 
${\cal J} \ll a_{\rm r}T^4$. This implies that the solution is not physical, and
moreover, the local thermodynamical equilibrium is not satisfied
although the relation ${\cal L}_{\mu\nu}={\cal J} h_{\mu\nu}/3$ is chosen for the
closure. This test is nothing but a numerical game.  Thus, even if a
numerical code can produce the stationary solution in this test, it
might not prove that the code is suitable for physical simulations.

\begin{table}[t]
 \caption{The chosen constants and initial data for the
one-dimensional test simulations.  Adiabatic index, $\Gamma$, an
artificially chosen value of $a_{\rm r}$, the value of $\kappa_0$,
$(\rho, P, u^x, {\cal J})$ for $x < 0$ and $x>0$, respectively.  
$a_{\rm r}$ is determined by ${\cal J}/(P/\rho)^4$ for $x>0$, and 
then, ${\cal J}$ for $x>0$ is determined by $a_{\rm r}(P/\rho)^4$. 
}
%%%%%%%%%%%%%%
\begin{center}
\begin{tabular}{cccc|c|c} \hline
No & $\Gamma$ & $a_{\rm r}$ & $\kappa_0$ & 
$(\rho, P ,u^x, {\cal J})_{x<0}$ & 
$(\rho, P ,u^x, {\cal J})_{x>0}$ 
\\  \hline \hline
1 & $5/3$ & $10^{10}/0.81$ & $0.4$ & 
(1.0, 3.0e-5, 0.015, 1.0e-8) &
(2.401, 1.612e-4, 0.006247, 2.509e-7) \\
2 & $5/3$ & $78125$ & $0.2$ & 
(1.0, 4.0e-3, 0.25, 2.0e-5) &
(3.109, 0.04512, 0.0804, 3.464e-3) \\
3 & $2$ & $10^7/6.48$ & $0.3$ & 
(1.0, 60, 10.0, 2.0) &
(7.9963, 1.25058, 2.342e3, 1.136e3) \\
4 & $5/3$ & $10^9/7.2$ & $0.08$ & 
(1.0, 6.0e-3, 0.69, 0.18) &
(3.65, 0.3588, 1.89, 1.297)\\ \hline
\end{tabular}
\end{center}
\label{table5}
\end{table}

%%%%%%%%%%%%%%%%%%%%%%%%%%%%%%%%%%%%%%%%%%%%%%%%%%%%%%%%%%%%%

\end{document}